\documentclass[reprint,superscriptaddress,nofootinbib]{revtex4-1}

\usepackage{graphicx}
\usepackage{dcolumn}
\usepackage{bm}
\usepackage{float}
\usepackage{amsthm,amssymb,amsmath,url,braket,amsbsy} %%%MATH
\usepackage{threeparttable}
\usepackage[utf8]{inputenc}
\usepackage{dirtytalk}
\graphicspath {{./figure_Qu_070823/}}
\usepackage{color}
\usepackage{cancel}
\newcommand\scalemath[2]{\scalebox{#1}{\mbox{\ensuremath{\displaystyle #2}}}}

\begin{document}

\title{Anisotropy of the spin Hall effect in a Dirac ferromagnet}

\author{Guanxiong Qu}
\affiliation{Department of Physics, The University of Tokyo, Bunkyo-ku, Tokyo 113-0033, Japan}
\affiliation{RIKEN Center for Emergent Matter Science (CEMS), Wako 351-0198, Japan.}
\author{Masamitsu Hayashi}
\affiliation{Department of Physics, The University of Tokyo, Bunkyo-ku, Tokyo 113-0033, Japan}
\affiliation{Trans-scale Quantum Science Institute (TSQSI), The University of Tokyo, Bunkyo-ku, Tokyo 113-0033, Japan}
\author{Masao Ogata}
\affiliation{Department of Physics, The University of Tokyo, Bunkyo-ku, Tokyo 113-0033, Japan}
\affiliation{Trans-scale quantum science institute (TSQSI), The University of Tokyo, Bunkyo-ku, Tokyo 113-0033, Japan}
\author{Junji Fujimoto}
\affiliation{Department of Physics, The University of Tokyo, Bunkyo-ku, Tokyo 113-0033, Japan}
%\affiliation{National Institute for Materials Science, Tsukuba 305-0047, Japan}

\date{\today}

\begin{abstract}
We study the intrinsic spin Hall effect of a Dirac Hamiltonian system with ferromagnetic exchange coupling, a minimal model combining relativistic spin-orbit interaction and ferromagnetism. 
The energy bands of the Dirac Hamiltonian are split after introducing a Stoner-type ferromagnetic ordering which breaks the spherical symmetry of pristine Dirac model. 
The totally antisymmetric spin Hall conductivity (SHC) tensor becomes axially anisotropic along the direction of external electric field.
Interestingly, the anisotropy does not vanish in the asymptotic limit of zero magnetization. 
We show that the ferromagnetic ordering breaks the spin degeneracy of the eigenfunctions and modifies the selection rules of the interband transitions for the intrinsic spin Hall effect.
The difference in the selection rule between the pristine and the ferromagnetic Dirac phases causes the anisotropy of the SHC, resulting in a discontinuity of the SHC as the magnetization, directed orthogonal to the electric field, is reduced to zero in the ferromagnetic Dirac phase and enters the pristine Dirac phase.
\end{abstract}

%\keywords{}

\maketitle

\date{\today}

\section{Introduction} \label{Sec1: Intro}

The spin Hall effect (SHE) allows generation of a spin current transverse to the external electric field within materials with large spin-orbit coupling (SOC)\cite{Sinova2015}. It is a standard method of spin current generation in the field of spintronics. The effect has been studied with strong interest in theoretical and experimental point of views \cite{Kato2004,Sinova2004,Murakami2006,Liu2012a}. In the early stage of research, substantial effort was put in understanding the SHE in paramagnetic materials with large SOC, e.g., heavy metals \cite{Tanaka_2008,Liu2012a,Miao2013}. Recent studies have expanded the system to include ferromagnetic and antiferromagnetic materials. First principles calculations \cite{Amin_2019,Qu_2020} have shown that the SHE in ferromagnets displays the interplay of ferromagnetic ordering and SOC. However, a general model that can account for such interplay has not been established.

One of the simplest models that includes both ferromagnetism and SOC is the Dirac ferromagnet \cite{fujimoto2014transport}. The SOC, a relativistic effect, naturally exists in the Dirac equation \cite{crepieux2001relativistic}. Despite its origin in relativistic particle physics, the Dirac equation is employed as a low energy effective model in condensed matter physics, partly because of its diverse mathematical forms \cite{thaller2013dirac}. For example, bismuth \cite{wolff1964matrix, fuseya2015transport} and the bulk states of three-dimensional topological insulators \cite{checkelsky2012dirac, PhysRevB.82.045122, zhang2009topological} are effectively described by the Dirac Hamiltonian based on $k \cdot p$ perturbation theory. Particularly, the Dirac Hamiltonian is equivalent with an isotropic Wolff Hamiltonian \cite{wolff1964matrix}, which effectively describes the low-energy states of semiconductors and semimetals with strong SOC and small gaps. Various transport properties that arise from such a unique Hamiltonian have been discussed \cite{fujimoto2014transport,fuseya2012spin,fukazawa2017intrinsic, PhysRevB.106.094414}. 

Ferromagnetic ordering can be incorporated into the Dirac equation through a Stoner-type mean field. Along this line, historical proposals were made: (i) a ferromagnetic order parameter having opposite signs between the positive and the negative energy states introduced by MacDonald and Vosko \cite{macdonald1979relativistic}; (ii) an order parameter having the same sign in both states proposed by Ramana and Rajagopal \cite{ramana1981theory}. It is known that ferromagnetic order can be induced in topological insulators by doping magnetic impurities \cite{checkelsky2012dirac, chang2013, yu2010quantized}. A recent study also showed the possibility of introducing ferromagnetism into a Dirac semimetal by a proximity effect \cite{PhysRevB.100.245148}.

It is well known that the two-dimensional Rashba-type Hamiltonian can be regarded as a minimal model to treat the anomalous Hall effect (AHE) in ferromagnets \cite{onoda2008quantum,Nagaosa2010,Ghosh2019PRB}. However, in the Rashba Hamiltonian, the form of SOC breaks the mirror symmetry and hence it cannot be used to study the coupling between ferromagnetism and SOC in mirror symmetric materials. In contrast, the Dirac Hamiltonian inherently includes the SOC without breaking its spherical symmetry, which is broken if ferromagnetic order is introduced in the system. We, thus, consider a system that can be described by a ferromagnetic Dirac Hamiltonian, i.e., a Dirac ferromagnet as a minimal model to study the SHE in the ferromagnets.

In this paper, we study the SHE of a Dirac ferromagnet with particular focus on its anisotropy. The intrinsic SHC is calculated using the Kubo formula in the clean limit. The SHC tensor is found to be axially anisotropic along the axis parallel to the external electric field. The anisotropy of the SHC tensor is found to be maintained even when the ferromagnetic ordering asymptotically vanishes. Comparing the eigenfunctions of the ferromagnet and pristine Dirac electrons through degenerate perturbation theory, the anisotropy of the SHC tensor and its asymptotic limits are explained by the modification of selection rules associated with the set of eigenfunctions in Dirac ferromagnet.

The paper is organized in the following order. In Sec.~\ref{Sec2: Model}, the theoretical model is defined, and the calculation method is presented.% in Sec.~\ref{Sec3: intrinsic SHC}. 
The main result is shown in Sec.~\ref{Sec3: intrinsic SHC} where the anisotropy of the intrinsic SHC tensor and its asymptotic behavior in zero magnetization limit are discussed. In Sec.~\ref{Sec5: Dirac vs Ferro}, we compare the ferromagnetic and pristine Dirac electron systems to explain the anisotropy and discontinuity of the SHC. A summary follows in Sec.~\ref{Sec6:Summary}.
%%%% End of Introduction %%%%%%%

%%%% Fig 1 %%%%%%%

%%%% end of Fig 1 %%%%%%%

%%%% Method %%%%%%%

\section{Model description} \label{Sec2: Model}

The model Hamiltonian of a Dirac ferromagnet is described by a $4\times4$ Dirac Hamiltonian with magnetization $\bm{M}=(M_1,M_2,M_3)$ representing the ferromagnetic ordering:
\begin{equation}
\mathcal{H}_0= \hbar v k_i \rho_1 \otimes \sigma^i +  \Delta \rho_3 \otimes \sigma^0 + M_i \rho_3 \otimes \sigma^i ,
\label{eq:2_1}
\end{equation}
where $\hbar$ is the reduced Planck's constant. $\rho_i$ $(i=1,2,3)$ and $\sigma^i$ $(i=1,2,3)$ are the Pauli matrices spanning the electron-hole space and the spin space, respectively.\footnote{$\bm{\rho}$ space is effectively the orbital space when taking the Dirac Hamiltonian as the effective Hamiltonian of multiorbital systems.} $\rho_0$ and $\sigma^0$ are $2 \times 2$ identity matrices.
The first two terms of the right hand side of Eq.~(\ref{eq:2_1}) are known to represent the low energy electronic states of systems with large SOC (and a small band gap), which is equivalent with an isotropic Wolff Hamiltonian  \cite{wolff1964matrix}. In the context of such a low-energy effective Hamiltonian, $v$ and $\Delta$ are the Fermi velocity and the band gap around the Dirac cone, respectively. See, for example, ref.~\cite{fuseya2012spin} in which the Hamiltonian is derived for the electronic states near the L point of bismuth. Note that the $ \bm{k} \cdot \bm{\sigma}$ term in Eq.~(\ref{eq:2_1}) contains the SOC, which originates from the off-diagonal components of velocity matrix in the $k\cdot p$ expansion. 

In Eq.~(\ref{eq:2_1}), we express the magnetization $\bm{M}$ using the spin magnetic moment operator $\rho_3 \otimes \sigma^i$ \cite{fuseya2015transport,PhysRevB.105.214419}, where the field $\bm{M}$ acts oppositely on electron (positive energy) and hole (negative energy) states \cite{macdonald1979relativistic}.  $M_i \rho_3 \otimes \sigma^i $ is the physical description of spontaneous magnetization since generators $\rho_3 \otimes \sigma_i$ also couple with the external magnetic field $\bm{B}$~\cite{Bruno_2001}. The generators $\rho_3 \otimes \sigma_i$ can be derived using the Peierls substitution and the Foldy and Wouthuysen transformation \cite{foldy1950dirac}. Details on the derivation of the Dirac Hamiltonian [Eq. (\ref{eq:2_1})] is shown in Appendix.~\ref{Appendix:0}. We do not consider another definition of ferromagnetic order ($S_i \rho_0 \otimes \sigma^i$) proposed in Ref.~\cite{ramana1981theory} in which the field $\bm{S}$ acts on electron and hole states in the same way. Additionally, we assume a weak magnetization limit that satisfies $M < \Delta $ so as to avoid the gap closing. Note that the case with $M > \Delta $ corresponds to a Weyl semimetal phase \cite{PhysRevB.84.235126, PhysRevB.93.045201,RevModPhys.90.015001}. In the following, we refer to the system represented by $\mathcal{H}_0$ with $\bm{M}=0$ ($\bm{M} \neq 0$) as the pristine (ferromagnetic) Dirac electron system. We choose $\hbar=1$ and renormalize the Fermi velocity with $v=1$. The Einstein notation is employed. 

\subsection{Rotation matrix}
Taking the unit direction of spontaneous magnetization $(\hat{m} \equiv \bm{M}/M)$ as a reference, the wave vector $\bm{k}$ can be separated into two parts: $k_{\parallel}$ represents the parallel component of $\bm{k}$ with $\hat{m}$ whereas $k_{\perp}$ is the perpendicular component, that is,
\begin{eqnarray}
\bm{k} = (\bm{k} \cdot \hat{m}) \hat{m} - (\bm{k} \times \hat{m}) \times \hat{m} \equiv k_{\parallel} \hat{m} + \bm{k}_{\perp}.
\label{eq:2_2}
\end{eqnarray}
Note that $k_{\perp}^2 = k^2 - (\bm{k} \cdot \hat{m})^2 $. In the spherical coordinate system, $\hat{m}$ is defined as: $\hat{m}=( \sin \theta \cos \varphi , \sin \theta \sin \varphi, \cos \theta )$. For simplicity, we define a rotation matrix $\mathcal{R}$, which rotates $\hat{m}$ to $\hat{z}$: $\mathcal{R} \hat{m} = \hat{z}$. The explicit form of the rotation matrix $\mathcal{R}$ is
\begin{equation}
\scalemath{0.9}{
\mathcal{R}=\left(\begin{array}{ccc} 
(1-\cos \theta) \cos^2 \varphi -1 & (1-\cos \theta) \sin \varphi \cos \varphi & \sin \theta \cos \varphi \\
 (1-\cos \theta) \sin \varphi \cos \varphi  & (1-\cos \theta) \sin^2 \varphi -1 & \sin \theta \sin \varphi \\
\sin \theta \cos \varphi & \sin \theta \sin \varphi & \cos \theta
\end{array}\right).}
\label{eq:2_4}
\end{equation}
Note that the inner product of $\bm{k}$ and $\bm{M}$ is $k_i M_i = \bm{k} \cdot \bm{M} = M \bm{k} \cdot \hat{m} = M k_{\parallel}$. Thus, the energy eigenstates of $\mathcal{H}_0$ read
\begin{eqnarray}
E_{\zeta,\eta} 
&=&  \zeta \sqrt{ k^2_{\parallel}+ (\sqrt{\Delta^2+k_{\perp}^2} + \eta M)^2},
\label{eq:2_5}
\end{eqnarray}
where $\zeta =\pm1$ represents the energy eigenstates [$\zeta =1$ for positive (electron) and $\zeta =-1$ for negative (hole) energy states], and $\eta =\pm1$ indicates the spin state ($\eta =1$ for spin-up and $\eta =-1$ spin-down) .

%%%%%%%%%%% fig:1 %%%%%%%%% 
 \begin{figure}[h]
 \includegraphics[width=8cm]{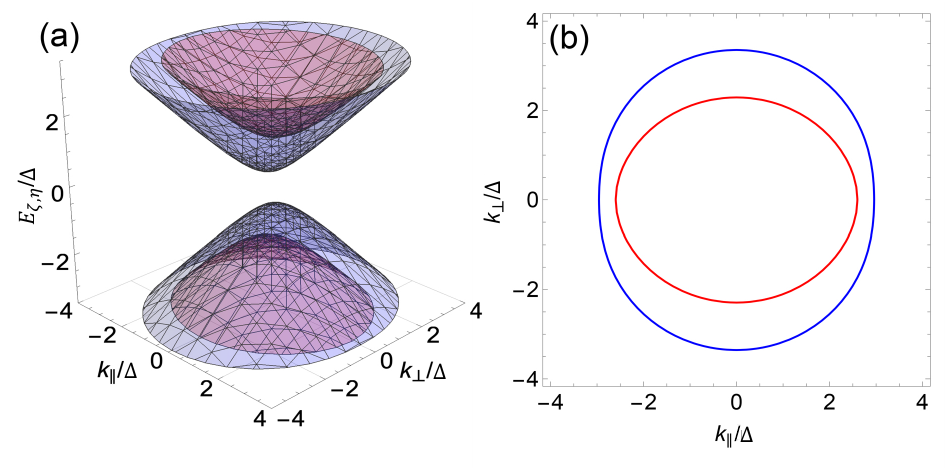}
 \caption{\label{fig:1}(a) Band dispersions and (b) Fermi contours ($\mu/\Delta=3$, positive energy branch) of the Dirac ferromagnet. The red colored bands correspond to $\eta=+1$, and blue colored bands correspond to $\eta=-1$. The magnetization is set to $M/\Delta=0.4$.}
 \end{figure}
 %%%%%%%%%%%  fig:1 %%%%%%%%% 
 
The band dispersion of pristine Dirac Hamiltonian ($\mathcal{H}_0$ with $\bm{M}=0$) is spherically symmetric, and the spin space is doubly degenerate. In the presence of magnetization, the spin degeneracy is broken, and spin-split bands appear as shown in Fig.~\ref{fig:1}(a). Note that the spin polarization of each spin state $(\eta=\pm1)$ is not necessarily parallel to the direction of $\bm{M}$, due to the inherent SOC of Dirac Hamiltonian. In addition, the Fermi surface becomes axially anisotropic along the direction of magnetization, which can be recognized by the Fermi contours on the $k_{\parallel}-k_{\perp}$ plane [see Fig.~\ref{fig:1}(b)].
 
\subsection{Green's function}
The electron Green's function $G^{(0)}(\varepsilon) = (\varepsilon - \mathcal{H}_0)^{-1}$ is rewritten with the generators of  $\rho_{\mu} \otimes \sigma^{\nu}$ \cite{fujimoto2014transport},
\begin{equation}
G^{(0)}(\varepsilon) = \frac{1}{D(\varepsilon)} g^{(0)}_{\mu\nu} \rho_{\mu} \otimes \sigma^{\nu},
\label{eq:2_6}
\end{equation}
where $\mu,\nu= 0,1,2,3$ and the denominator $D(\varepsilon)=\prod\limits_{\eta,\zeta= \pm 1}  (\varepsilon -   E_{\eta,\zeta})$. The 16 components of the numerator $g^{(0)}_{\mu\nu}$ are listed in Table~\ref{Tab:1}.
Here, we set $\varepsilon_{\bm{k}}= \sqrt{k^2+\Delta^2 -M^2}$.

\begin{table*}
\caption{\label{Tab:1} Expression of $g_{\mu \nu}^{(0)}$ in the Green's function of  the Dirac ferromagnet. Indices $\mu,\nu$ are listed in columns and rows.}
\begin{ruledtabular}
\begin{tabular}{ccccc}
$(\mu,\nu)$ & 0 & 1 & 2 & 3 \\ 
\hline
0 & 
$ \varepsilon ( \varepsilon^2 - \varepsilon_{\bm{k}}^2) -2 \varepsilon M^2$ & 
$-2 \Delta \varepsilon M_1$& 
$-2 \Delta \varepsilon M_2$&
$-2 \Delta \varepsilon M_3 $\\
1 & 
$-2 \Delta Mk_{\parallel}$&
$k_1 (\varepsilon^2 - \varepsilon_{\bm{k}}^2 )  -2 M_1 Mk_{\parallel} $&
$k_2 (\varepsilon^2 - \varepsilon_{\bm{k}}^2 )  -2 M_2 Mk_{\parallel} $ &
$k_3  (\varepsilon^2 - \varepsilon_{\bm{k}}^2 ) -2 M_3 Mk_{\parallel} $\\
2 &
0 &
$-2\varepsilon (k_2 M_3 - k_3 M_2) $& 
$-2\varepsilon( k_3 M_1 - k_1 M_3)$&
$-2\varepsilon( k_1 M_2 - k_2 M_1)$ \\
3 &
$ \Delta (\varepsilon^2 - \varepsilon_{\bm{k}}^2 )$ &
$-  M_1(\varepsilon^2 + \varepsilon_{\bm{k}}^2) +2 Mk_{\parallel}$ &
$-  M_2(\varepsilon^2 + \varepsilon_{\bm{k}}^2)  +2 Mk_{\parallel}$ &
$-  M_3(\varepsilon^2 + \varepsilon_{\bm{k}}^2) +2 Mk_{\parallel}$  \\
\end{tabular}
\end{ruledtabular}
\end{table*}

\subsection{Charge and spin velocity operator}

The velocity operator is obtained directly from the Dirac Hamiltonian Eq.~(\ref{eq:2_1}),
\begin{eqnarray}
v_{i} &=& \frac{ \partial \mathcal{H}_0}{ \partial k_i} = \rho_1 \otimes \sigma^{i},
\label{eq:2_7}
\end{eqnarray}
where the charge current operator is defined as $j_{i}=-e v_{i}$. $i=1,2,3 (x,y,z)$ represents the Cartesian coordinates of the velocity.

The spin velocity operator is given by the anticommutator of the velocity operator and the spin operator $s^{k}= \rho_3 \otimes \sigma^{k}$ \cite{fuseya2012spin,fukazawa2017intrinsic}, that is,
\begin{eqnarray}
v_{i}^{k }&=& \frac{1}{2} \{ v_{i},  s^{k} \} = \epsilon_{i k j} 
\rho_2 \otimes \sigma^{j},
\label{eq:2_8}
\end{eqnarray}
where $ \epsilon_{i k j} $ is the Levi-Civita symbol and $k= 1 ,2 ,3 (x,y,z)$ represents the Cartesian coordinates of the spin direction.
We define the spin current operator as $j^{k}_{i}= \frac{\hbar}{2} v_{i}^{k}$.

%%%%%%% Model Hamiltonian %%%%%%%%

\section{Intrinsic spin Hall conductivity}\label{Sec3: intrinsic SHC}
The intrinsic SHC is calculated through the Kubo formula. The correlation function between charge current $j_j$ and spin current $j^k_i$ reads
\begin{eqnarray}
Q_{ij}^{k} (i v) &=& - \frac{1}{V} \int_{0}^{\beta} du \ e^{iv u} \braket{ \hat{T} j_i^{k}(u) j_j(0)}, \nonumber \\
&=& \frac{1}{V \beta} \sum_{\bm{k},n} tr\left[ \tilde{G}^{(0)}(i \omega_n) j^{k}_{i}   \tilde{G}^{(0)} (i \omega_n+iv) j_{j} \right] ,
\label{eq:3_1}
\end{eqnarray}
where $iv \rightarrow \omega + i 0$ is analytic continuation of the response frequency $\omega$ with Matsubara frequency $v=2m\pi/\beta$ and $\beta$ is the inverse temperature. $\tilde{G}^{(0)}(i \omega_n)$ is the electron Green's function with Matsubara frequency $w_n=(2n+1)\pi/\beta$ and chemical potential $\mu$,
\begin{equation}
\tilde{G}^{(0)}(i \omega_n) = \frac{1}{D(i \omega_n+\mu)} g^{(0)}_{\mu\nu}(i \omega_n+\mu) \rho_{\mu}  \otimes \sigma^{\nu}.
\label{eq:3_2}
\end{equation}
The intrinsic SHC is obtained by taking the static limit of the correlation function $Q_{ij}^{k} (i v)$,
\begin{eqnarray}
\sigma_{ij}^{k} =   \lim_{v \to 0} \frac{Q_{ij}^{k} (i v) -Q_{ij}^{k} (0)}{-v}.
\label{eq:3_3}
\end{eqnarray}

%\begin{widetext}
Assuming zero temperature $(\beta \to \infty)$ and evaluating the Matsubara summation, the intrinsic SHC can be separated into two parts $\sigma_{ij}^{k} = \sigma_{ij}^{k,(1)} + \sigma_{ij}^{k,(2)}$ with
\begin{align}
 \label{eq:3_4}
\sigma_{ij}^{k,(1)} &=   \frac{e}{8 \pi V} \sum_{\bm{k}}   tr\Big[  - 2G^{A}(\varepsilon)  v^{k}_{i}  G^{R}(\varepsilon)  v_j  \notag \\
& +  G^{R}(\varepsilon)  v^{k}_{i}  G^{R}(\varepsilon)  v_j  + G^{A}(\varepsilon)  v^{k}_{i}  G^{A}(\varepsilon) v_j  \Big]   \Big|_{\varepsilon=\mu}  , \\
 \label{eq:3_5}
 \sigma_{ij}^{k,(2)} &=  \frac{e}{8 \pi V} \sum_{\bm{k}} \int^{\mu}_{-\infty} d\varepsilon \ tr\Big[ G^{R}(\varepsilon)  v^{k}_{i}  \partial_\varepsilon G^{R}(\varepsilon)  v_j \notag\\
 &-   \partial_\varepsilon G^{R}(\varepsilon)  v^{k}_{i} G^{R}(\varepsilon) v_j   - \Big( R \leftrightarrow A\Big)   \Big] ,
\end{align}
where $G^{R,A}(\varepsilon)= G^{(0)}(\varepsilon\pm i \gamma)$ are the retarded and advanced Green's function with a damping constant $\gamma$. 
%\end{widetext
The detailed structure of the self-energy (damping constant) in the Dirac Hamiltonian has been considered using short-range impurities~\cite{fujimoto2014transport,fukazawa2017intrinsic,PhysRevB.106.094414}. Here we focus on the intrinsic contribution to the SHE where we first introduce a damping constant $\gamma$ and later take it to the clean limit $\gamma \to 0$. The extrinsic contribution can be obtained by extending the calculation on self-energy $\gamma$ or by employing semiclassical Boltzmann equation \cite{Sinitsyn_2007} to include the effect of disorder. The intrinsic contribution is dissipationless and therefore it cannot be treated by the Boltzmann equation approach. In the zero temperature assumption, the integral in $\sigma_{ij}^{k,(1)}$ reduces to a surface term $\varepsilon=\mu$, whereas, $\sigma_{ij}^{k,(2)}$ contains integration over energy up to the chemical potential $\mu$. Consequently, $\sigma_{ij}^{k,(1)}$  is often referred to as the ``Fermi-surface" term, and $ \sigma_{ij}^{k,(2)}$ is known as the ``Fermi sea" term \cite{Streda_1982}.

Here, we take one of the nonvanishing tensor components $\sigma^{3}_{21}$ as an example, in which, the direction of magnetization is rotated arbitrarily. Through integration by parts, the Fermi-surface term exactly cancels part of the Fermi sea term. See Appendix.~\ref{Appendix:A}.3. The total intrinsic SHC, thus, reduces to a compact form,
\begin{align}
\sigma_{21}^{3} &=\sigma_{21}^{3,(1)} + \sigma_{21}^{3,(2)} \notag \\
&=
 \frac{ 2e }{  V} \sum_{\bm{k}}      \int^{ \mu}_{-\infty} d\varepsilon\  \text{sgn} D'( \zeta E_{\eta}) \delta(D(\varepsilon))  \partial_\varepsilon \Big(   \frac{X^{(0)}(\varepsilon) }{D'(\varepsilon)}  \Big),
\label{eq:3_18}
\end{align}
where $ X^{(0)}(\varepsilon) = s_{\alpha}X_{\alpha}^{(0)}(\varepsilon)$ and $X_{\alpha}^{(0)}(\varepsilon)$ is defined as
\begin{eqnarray}
X_{\alpha}^{(0)}(\varepsilon) \equiv  g^{(0)}_{0 \alpha }(\varepsilon)\partial_\varepsilon  g^{(0)}_{3 \alpha}(\varepsilon) - \partial_\varepsilon g^{(0)}_{0 \alpha}(\varepsilon)  g^{(0)}_{3 \alpha}(\varepsilon).
 \label{eq:3_12}
\end{eqnarray}
Equation~(\ref{eq:3_18}) clearly shows that the intrinsic SHE in the Dirac ferromagnet is purely a Fermi sea effect. This is analogous to the intrinsic AHE: the anomalous Hall conductivity (AHC) is obtained by integrating the Berry curvature of all bands below the Fermi surface, corresponding to an interband mixing effect independent of the relaxation time \cite{Haldane_2004,Sinitsyn_2007}. Through a straightforward calculation, $\sigma_{21}^{3}$ can be separated into two terms with respect to $\hat{m}$,
\begin{align}
 \label{eq:4_4}
\sigma_{21}^{3}
&\equiv \sigma^{3,\mathrm{iso}}_{21} +m_1^2 (\sigma^{3,\hat{m}}_{21}-\sigma^{3,\mathrm{iso}}_{21}),
\end{align}
where we define $\sigma^{3,\mathrm{iso}}_{21}$ and $\sigma^{3,\hat{m}}_{21}$ as
\begin{align}
 \label{eq:4_5}
\sigma^{3,\mathrm{iso}}_{21} &\equiv - \frac{ e \Delta }{ 4 V}\sum_{\bm{k},\eta,\zeta}  \zeta  \int^{ \mu}_{-\infty} d\varepsilon\     \delta(\varepsilon- \zeta E_{\eta}) \notag \\
& \times  \left( \frac{  1 }{ {E_{\eta}}  \tilde{k}_{\perp}^2 }     +  \frac{ \eta  }{ M } \frac{1 }{ {E_{\eta}}  \tilde{k}_{\perp} }   -   \frac{ \eta  }{ M }  \frac{ {E_{\eta}}}{   \tilde{k}_{\perp}^3 }  \right) ,\\
 \label{eq:4_6}
\sigma^{3,\hat{m}}_{21} &\equiv -\frac{ e \Delta }{ 4 V}\sum_{\bm{k},\eta,\zeta}  \zeta \int^{ \mu}_{-\infty} d\varepsilon\     \delta(\varepsilon- \zeta E_{\eta}) \notag \\
& \times \left( - \frac{  \eta  M   }{ E_{\eta}^3   \tilde{k}_{\perp} }  -   \frac{ 1}{ E_{\eta}^3  } \right),
\end{align}
and $ \tilde{k}_{\perp} = \sqrt{ k^2_{\perp} +\Delta^2 }$.

Equations~(\ref{eq:4_4})--(\ref{eq:4_6}) are the main results of this paper. The forms clearly show the symmetry of the intrinsic SHC in Dirac ferromagnet. $\sigma^{3,\mathrm{iso}}_{21}$ only depends on the strength of magnetization ($M$), which is the isotropic contribution to $\sigma^{3}_{21}$. On the contrary, the term $\sigma^{3,\hat{m}}_{21}-\sigma^{3,\mathrm{iso}}_{21}$ depends on the strength and the direction of the magnetization. The latter is the source of an anisotropic contribution to $\sigma^{3}_{21}$. Surprisingly, the intrinsic SHC is anisotropic along the $x$ axis, which is parallel to the external electric-field $\hat{E}$ (charge current). 
%instead of the direction of spin current polarization, i.e., $z$-axis. 
Note that we take $\sigma^3_{21}$ as an example where the external electric field is applied along the $x$ axis and the polarization of spin current is parallel to the $z$ axis. $\sigma^{3,\mathrm{iso}}_{21}$ and $\sigma^{3,\hat{m}}_{21}$ both contain $\varepsilon$ integrals with an integration limit that approaches minus infinity $(\varepsilon \to - \infty)$. One needs to introduce a proper energy cutoff $(\Lambda_{\varepsilon})$ to avoid the ultraviolet divergence in the effective Dirac Hamiltonian \cite{fujimoto2013ultraviolet,takane2019gauge}. See Appendix~\ref{Appendix:A}.5 for the details. 

We first present the results when the Fermi level lies within gap of the Dirac cone ($\mu=0$). The SHC reads
\begin{align}
\sigma^3_{21} = - \frac{ e \Delta }{ 4\pi ^2} \left[  m_1^2 (\ln  \frac{2 \Lambda_{\varepsilon}}{\Delta}  -1) - \frac{M^2}{ 6 \Delta^2} (1-4 m_1^2) \right] +\mathcal{O} (\tilde{M}^4). \notag \\
\label{eq:4_18}
\end{align}
As $M$ approaches zero, we may drop the second-order term $\mathcal{O}(\tilde{M}^2)$.
Interestingly, the zeroth-order term, independent of $M$, is dependent on the direction of magnetization, i.e., $m_1$. Consequently, the value of SHC differs when $\bm{M}$ is reduced to zero from the ferromagnetic state along different directions. For example, if the magnetization is oriented along the $x$ axis $(m_1=1)$, $\sigma^3_{21}$ takes a nonzero value when $M$ is reduced to zero, whereas $\sigma^3_{21}$ vanishes with $\bm{M} \rightarrow 0$ when the magnetization is orthogonal to the $x$ axis $(m_1=0)$.
Note that $\sigma^3_{21}$ takes a nonzero value for the pristine Dirac phase ($M=0$) \cite{fuseya2012spin,fukazawa2017intrinsic}.
Thus, the $M \rightarrow 0$ limit of the ferromagnetic Dirac phase does not match that of the pristine Dirac phase when the magnetization of the former is orthogonal to the $x$ axis.

%%%%%%%%%% fig:2 %%%%%%%%% 
 \begin{figure}[h]
 \includegraphics[width=8cm]{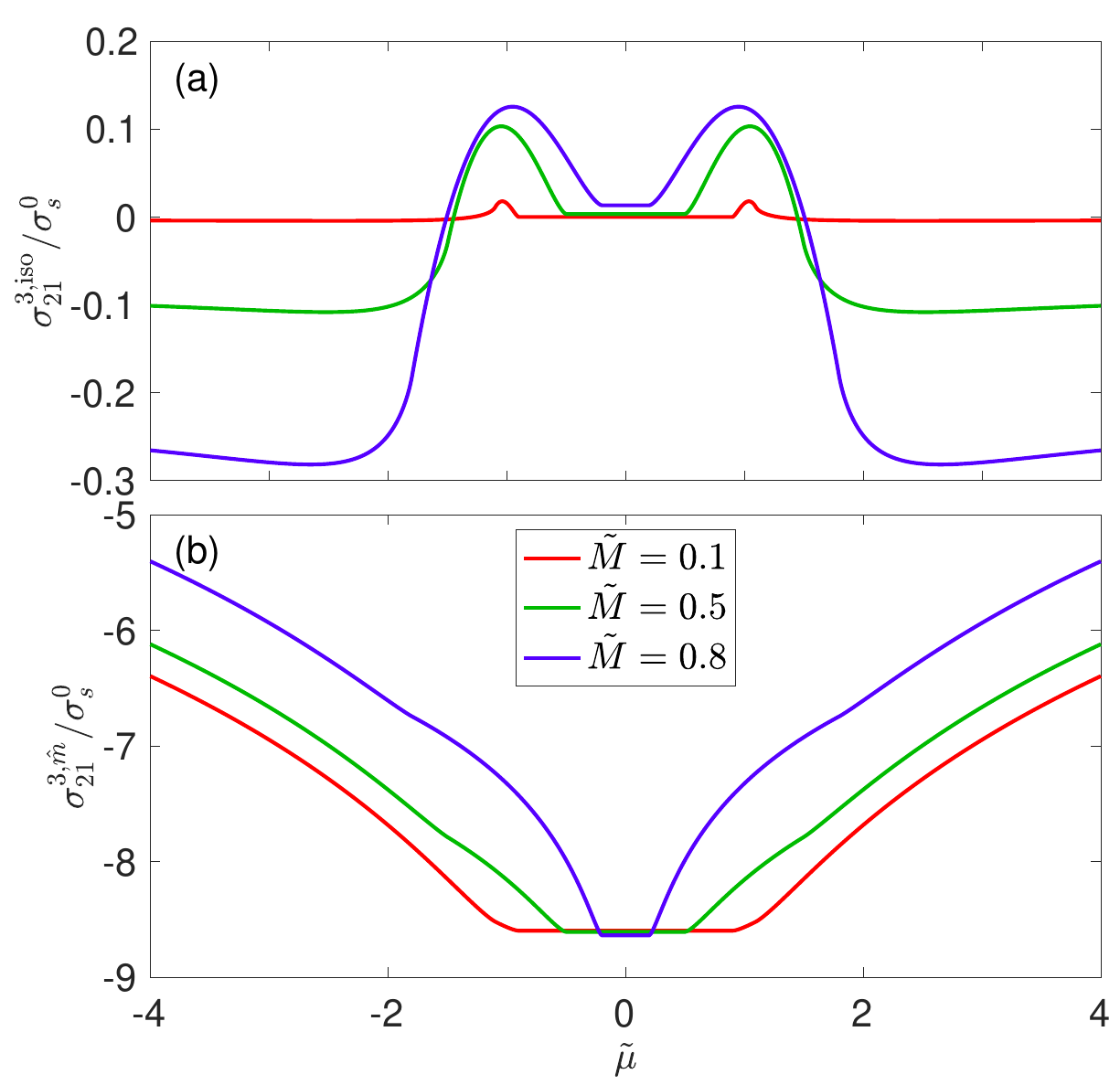}
 \caption{\label{fig:2} Chemical potential dependence of intrinsic SHC contributions: (a) $\sigma^{3,\mathrm{iso}}_{21}$ and (b) $\sigma^{3,\hat{m}}_{21}$. The strength of magnetization $\tilde{M}$ is set to $0.1, 0.5, 0.8$. The energy cutoff is set to $\Lambda_\varepsilon=100$. The conductivity is in the unit of $\sigma^0_{s}=e \Delta /8\pi^2$.  The chemical potential is normalized by $\Delta $, i.e., $\tilde{\mu} \equiv \mu/\Delta $. }
 \end{figure}
 %%%%%%%%%%%  fig:2 %%%%%%%%% 
 
In Fig.~\ref{fig:2}, we present the chemical potential dependence of the isotropic contribution $\sigma^{3,\mathrm{iso}}_{21}$ and the anisotropic contribution $\sigma^{3,\hat{m}}_{21}$. 
The intrinsic SHC is even with respect to $\mu$.
This is in contrast to the intrinsic AHC in the Dirac ferromagnet which is known to be odd with $\mu$ \cite{fujimoto2014transport}. 
Both $\sigma^{3,\mathrm{iso}}_{21}$ and $\sigma^{3,\hat{m}}_{21}$ have plateaus in the band gap ($|\mu / \Delta| \lesssim 1$).
The width of the plateau linearly reduces with increasing $M$.
The sign of $\sigma^{3,\mathrm{iso}}_{21}$ and $\sigma^{3,\hat{m}}_{21}$ at the plateau is opposite.
When the chemical potential is placed near the edge of the band gap, $\sigma^{3,\mathrm{iso}}_{21}$ increases compared to that at the plateau.
$\sigma^{3,\mathrm{iso}}_{21}$ takes a maximum at $\mu= \Delta$ for all strengths of $M$.
A further increase in $|\mu|$ causes $\sigma^{3,\mathrm{iso}}_{21}$ to decay and change its sign.
In contrast, $\sigma^{3,\hat{m}}_{21}$ monotonically decreases with increasing $|\mu|$. 
Overall, the magnitude of $\sigma^{3,\hat{m}}_{21}$ is significantly larger than that of $\sigma^{3,\mathrm{iso}}_{21}$.
Thus, the intrinsic SHC of the Dirac ferromagnet is dominated by the $\hat{m}$-dependent term ($\sigma^{3,\hat{m}}_{21}$).
In the following, we discuss the anisotropy of $\sigma_{21}^3$ in detail.

\subsection{The anisotropy of intrinsic SHC}
\begin{figure}[h]
 \includegraphics[width=8cm]{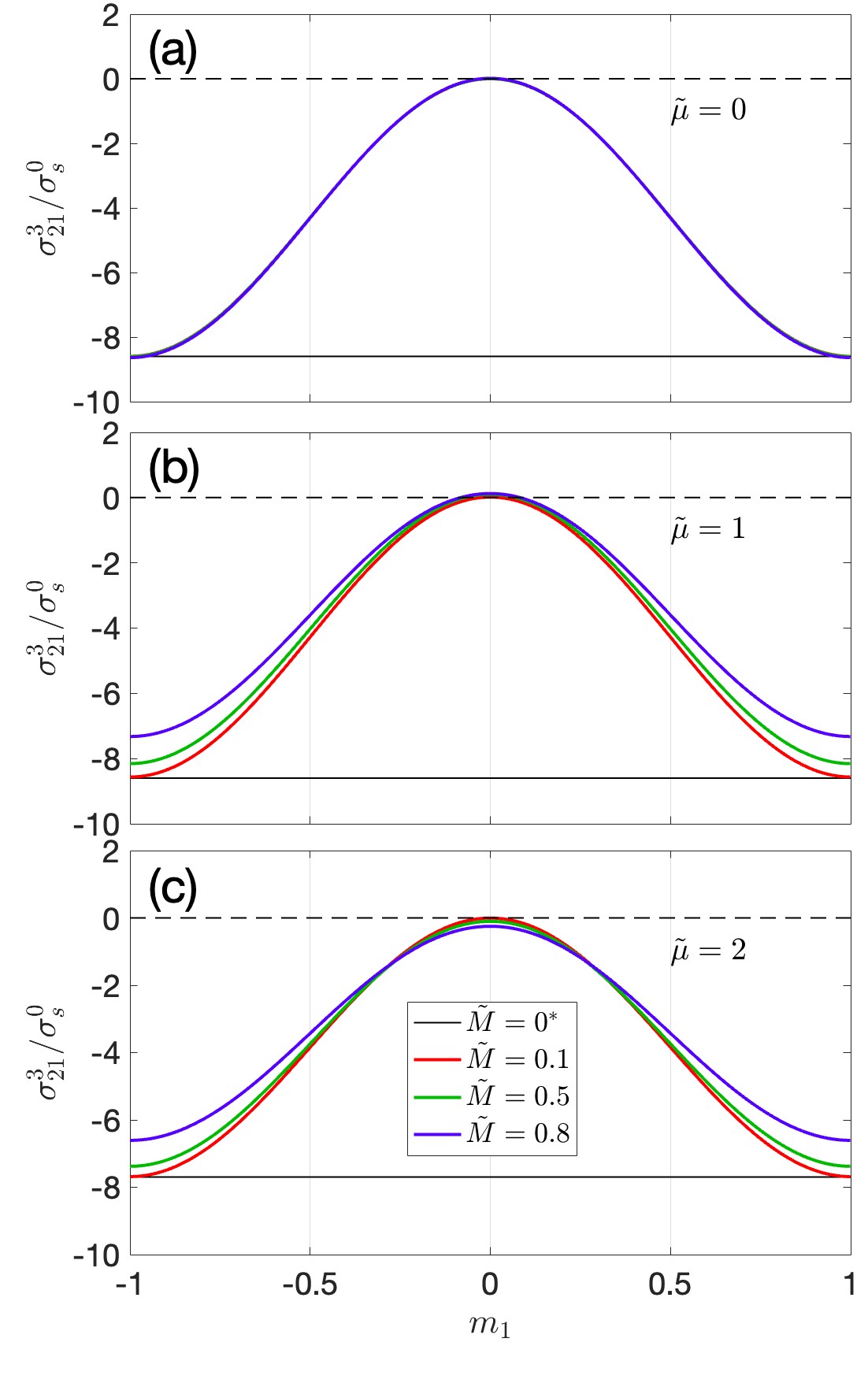}
 \caption{\label{fig:3} The angular dependence $(m_1)$ of intrinsic SHC $\sigma^{3}_{21}$ for various chemical potentials: (a) $\tilde{\mu} \equiv \mu/\Delta =0$, (b) $\tilde{\mu}=1$, and (c) $\tilde{\mu}=2$. The black solid line (*) shows the results for the pristine Dirac Hamiltonian \cite{PhysRevB.105.214419}. The strength of the magnetization $\tilde{M}\equiv M/ \Delta $ is set to $0.1, 0.5, 0.8$. The energy cutoff  is set to $\Lambda_\varepsilon=100$. The conductivity is in the unit of $\sigma^0_{s}=e \Delta  /8\pi^2$.
 }
 \end{figure}

\begin{figure}[h]
 \includegraphics[width=8cm]{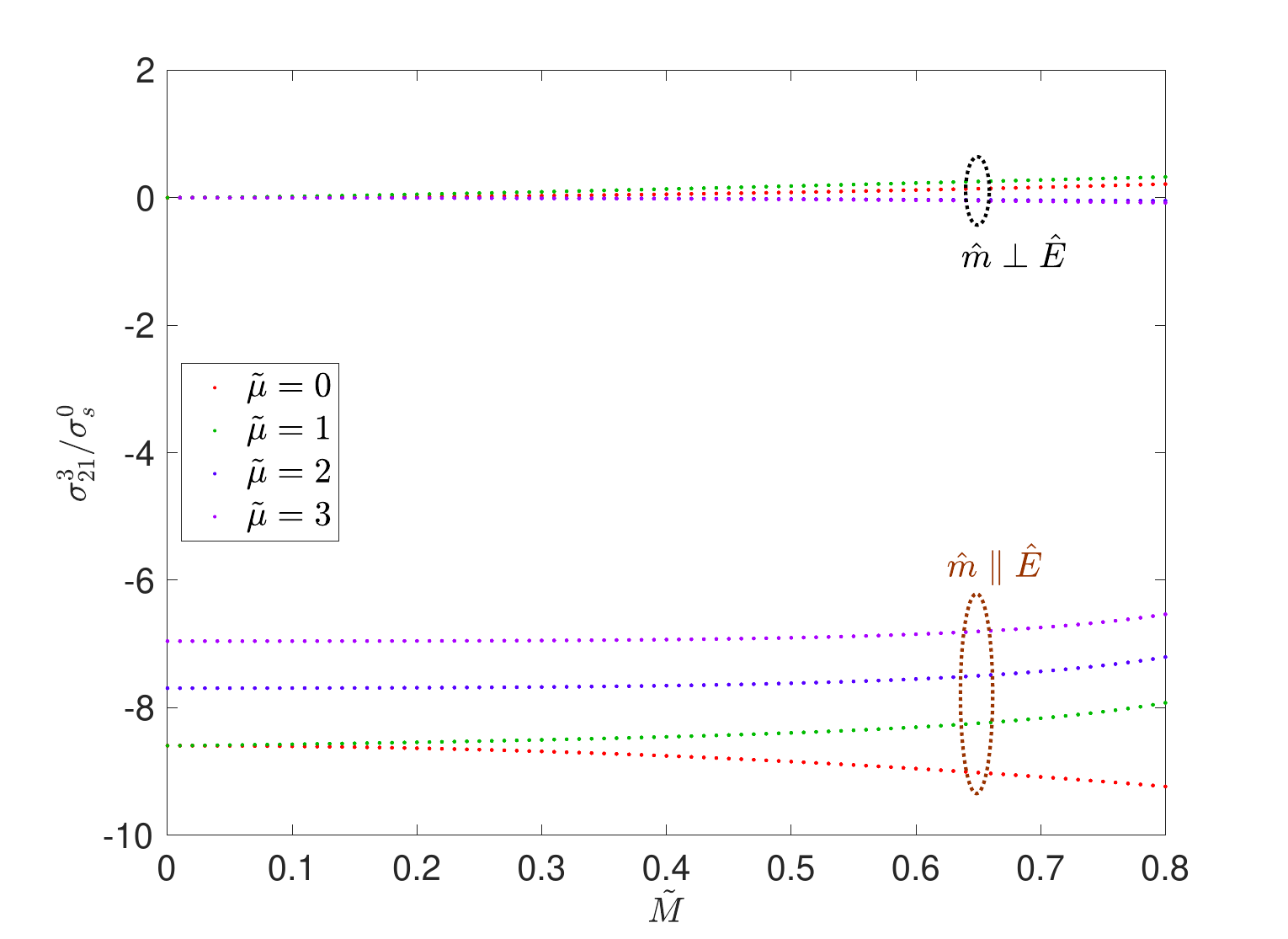}
 \caption{\label{fig:4} The magnetization strength dependence $\tilde{M}$ of intrinsic SHC $\sigma^3_{21}$ for various chemical potential $\tilde{\mu}$'s. The circled line groups correspond to the magnetization direction parallel with or perpendicular to $\hat{x}$. The energy cutoff  is set to $\Lambda_\varepsilon=100$. The conductivity is in the unit of $\sigma^0_{s}=e \Delta  /8\pi^2$.
 }
 \end{figure}

 $\sigma^3_{21}$  is calculated and plotted as a function of magnetization direction in Fig.~\ref{fig:3} where the strength of the magnetization and the position of the chemical potential are varied.
$|\sigma^3_{21}|$ takes a maximum when $\bm{M}\parallel \hat{x}$ $(m_1=\pm1)$ and is nearly zero when $\bm{M}\perp \hat{x}$ $(m_1=0)$.
%\textcolor{red}{(Shouldn't we plot $\sigma^3_{21}$ as a function of the magnetization angle if we are going to discuss anisotropy?)}
The SHC, thus, shows a strong uniaxial anisotropy where its symmetry axis is aligned along the direction of the electric field (charge current).
The anisotropy is not strongly influenced by the position of the chemical potential and the strength of magnetization.
%$\sigma^3_{21}$ exhibits a strong Specifically, 
For the latter, $\sigma^3_{21}$ is independent of $M$ when the chemical potential is inside the band gap $(\mu=0)$, see Fig.~\ref{fig:3}(a).
%%%%%%%%%% fig:3 %%%%%%%%% 
 %%%%%%%%%%%  fig:3 %%%%%%%%% 
Strikingly, the anisotropy of $\sigma^3_{21}$ does not vanish when $M$ asymptotically approaches zero. That is, $\sigma^3_{21}$ takes a nonzero value when $M$ is reduced to zero along the direction of external electrical-field $\hat{E}$, whereas, $\sigma^3_{21}$ vanishes when $M$ approaches zero with its direction set perpendicular to $\hat{E}$ [see Fig.~\ref{fig:4}]. In comparison, the AHE in the Dirac ferromagnet asymptotically vanishes with the strength of magnetization \cite{fujimoto2014transport} regardless of the magnetization direction. These results, thus, suggest that the Dirac ferromagnet model is not smoothly connected with the pristine Dirac model when ferromagnetic ordering vanishes asymptotically. This discrepancy is discussed in the following section.

%%%%%%%%%% fig:4 %%%%%%%%% 
 %%%%%%%%%%%  fig:4 %%%%%%%%% 

%%%%%%%%%%%%%%%%% Part II
\section{Pristine VS ferromagnetic Dirac electron systems  \label{Sec5: Dirac vs Ferro}}
To account for the asymptotic behavior of the intrinsic SHC in the Dirac ferromagnet, we compare the Hamiltonians of ferromagnetic and pristine Dirac systems.
The former is given in Eq.~(\ref{eq:2_1}), whereas, the latter is defined by
\begin{equation}
\mathcal{H}_D= \hbar v k_i \rho_1 \otimes \sigma^i + \Delta \rho_3 \otimes \sigma_0.
\label{eq:5_1}
\end{equation}
The eigenfunctions read
\begin{align}
\Psi^D_{\zeta,\eta} = \frac{1}{\sqrt{2}} \left(\begin{array}{c} \sqrt{1 + \zeta \frac{ \Delta }{ \tilde{k}}} \zeta \frac{ \bm{k} \cdot \bm{\sigma} }{|\bm{k}|}\chi_{\eta} \\ \sqrt{1 - \zeta  \frac{ \Delta  }{\tilde{k}}  } \chi_{\eta} \end{array}\right),
\label{eq:5_2}
\end{align}
where $\tilde{k}=\sqrt{\Delta^2 +k^2}$ and spinor $\chi_{\eta}$ is defined as $\chi_{+1}= (1\ 0)^T$, $\chi_{-1}= (0\ 1)^T$.  Again, we choose $\hbar=v=1$ for simplicity.

For the Dirac ferromagnet, here we use a specific magnetization direction $\bm{M}=(0,0,M)$ for simplicity. The specific direction of magnetization can be rotated to arbitrary direction by a global unitary transformation without loss of generality (see Appendix~\ref{Appendix:B} for the details).
The ferromagnetic Dirac Hamiltonian [Eq.~(\ref{eq:2_1})] is, then, simplified to
\begin{equation}
\mathcal{H}_F= \mathcal{H}_D + \mathcal{H}'(M),
\label{eq:5_3}
\end{equation}
where the magnetization-dependent part of the Hamiltonian is $ \mathcal{H}'(M)= M \rho_3 \otimes \sigma^3$. The corresponding eigenfunctions are
 \begin{align}
\Psi^F_{\zeta,\eta} = \frac{1}{2}\left(\begin{array}{c} 
\zeta e^{-i \phi} \sqrt{1+ \eta \frac{ \Delta  }{\tilde{k}_{\perp} } } \sqrt{1+ \zeta \eta \frac{Q_{\eta} }{E_{\eta}} } \\ 
- \eta  \sqrt{1- \eta \frac{ \Delta }{\tilde{k}_{\perp} } } \sqrt{1- \zeta \eta \frac{Q_{\eta} }{E_{\eta}} }  \\
e^{-i \phi}\sqrt{1+ \eta \frac{ \Delta }{\tilde{k}_{\perp} } } \sqrt{1- \zeta \eta \frac{Q_{\eta} }{E_{\eta}} }  \\ 
 \zeta \eta   \sqrt{1- \eta \frac{ \Delta  }{\tilde{k}_{\perp} } } \sqrt{1+ \zeta \eta \frac{Q_{\eta} }{E_{\eta}} } \end{array}\right),
\label{eq:5_4}
\end{align}
where $k_1=k_{\perp} \cos \phi$, $k_2=k_{\perp} \sin \phi$, $k_3=k_{\parallel}$, $Q_{\eta}= \tilde{k}_{\perp} + \eta M$, and $E_{\eta}=\sqrt{ k^2_{\parallel}+(\tilde{k}_{\perp} + \eta M)^2}$. 
As evident, $\Psi^D_{\zeta,\eta}$ and $\Psi^F_{\zeta,\eta}$ do not take the same form in the paramagnetic limit ($M \to 0$).
%, i.e., $\Psi^D_{\zeta,\eta}  \neq \Psi^F_{\zeta,\eta} \Big|_{M \to 0} $. 
This discrepancy causes the difference in $\sigma^3_{21}$ for pristine and ferromagnetic Dirac electron systems when $M$ for the latter asymptotically approaches zero along the direction of the external electrical field.

\subsection{Degenerate perturbation theory}
To trace the discrepancy between $\Psi^D_{\zeta,\eta}$ and $\Psi^F_{\zeta,\eta}$ in the paramagnetic limit, we construct the eigenfunctions of the Dirac ferromagnet from $\Psi^D_{\zeta,\eta}$ using degenerate perturbation theory \cite{sakurai1995modern}.
Here the magnetization $\mathcal{H}'$ is treated as perturbation.

First, we define the following projection operator that maps the total Hamiltonian to each spin-degenerate energy subspace $(\zeta = \pm 1)$ as
%$(\zeta= \pm 1)$ is
\begin{align}
P_{\zeta} &= \sum_{\eta} \Psi^D_{\zeta,\eta} \Psi^{D\dagger}_{\zeta,\eta} = \frac{1}{2} \left(\begin{array}{cc}  (1+\zeta \frac{ \Delta  }{ \tilde{k} })  I_2 & \zeta \frac{ \bm{k} \cdot \bm{\sigma} }{\tilde{k}}  \\ \zeta \frac{ \bm{k} \cdot \bm{\sigma} }{\tilde{k}}  &   (1-\zeta \frac{ \Delta  }{ \tilde{k} })  I_2 \end{array}\right).
\label{eq:5_5}
\end{align}
The projected Hamiltonian $\mathcal{H}_{\zeta}$ in each energy subspace reads
\begin{align}
\mathcal{H}_{\zeta} &=P_{\zeta} (\mathcal{H}_D + \mathcal{H}') P_{\zeta} .
\label{eq:5_6}
\end{align}
The unperturbed Hamiltonian $\mathcal{H}_D$ is naturally diagonalized in each energy subspace $\mathbb{H}_{\zeta}$, whereas, the magnetization $ \mathcal{H}'$ breaks the spin degeneracy. Hence, a new basis set $ \Phi^{(0)}_{\zeta,\eta}$ is required to diagonalize the perturbation $P_{\zeta}  \mathcal{H}' P_{\zeta} $ to the first order,
\begin{align}
P_{\zeta}   \hat{\mathcal{H}}'  P_{\zeta}  \Phi^{(0)}_{\zeta,\eta} = E^{(1)}_{\zeta,\eta} \Phi^{(0)}_{\zeta,\eta},
\label{eq:5_7}
\end{align}
where $E^{(1)}_{\zeta,\eta} = \zeta \eta M   \tilde{k}_{\perp} /\tilde{k}  $ is the first order correction of the eigenenergy and $\Phi^{(0)}_{\zeta,\eta}$ explicitly reads
\begin{align}
\Phi^{(0)}_{\zeta,\eta} = \frac{1}{2} \left(\begin{array}{c} 
\zeta e^{-i \phi} \sqrt{1+ \eta \frac{ \Delta  }{\tilde{k}_{\perp} } } \sqrt{1+ \zeta \eta \frac{\tilde{k}_{\perp} }{ \tilde{k} } } \\ 
- \eta  \sqrt{1- \eta \frac{ \Delta  }{\tilde{k}_{\perp} } } \sqrt{1- \zeta \eta \frac{\tilde{k}_{\perp} }{ \tilde{k} } }  \\
e^{-i \phi}\sqrt{1+ \eta \frac{ \Delta  }{\tilde{k}_{\perp} } } \sqrt{1- \zeta \eta \frac{\tilde{k}_{\perp} }{ \tilde{k} } }  \\ 
 \zeta \eta   \sqrt{1- \eta \frac{ \Delta  }{\tilde{k}_{\perp} } } \sqrt{1+ \zeta \eta\frac{\tilde{k}_{\perp} }{ \tilde{k} } } \end{array}\right)
\label{eq:5_8}
\end{align}
which is the zeroth order $(\mathcal{O}(M^0))$ correction of the eigenfunction.
Apparently, $\Phi^{(0)}_{\zeta,\eta}$'s are the eigenfunctions of the Dirac ferromagnet in the paramagnetic limit: $\Phi^{(0)}_{\zeta,\eta}= \Psi^F_{\zeta,\eta} |_{M \to 0}$. 
$ \Phi^{(0)}_{\zeta,\eta}$ and $\Psi^D_{\zeta,\eta}$ span the same spin-degenerate space, but $ \Phi^{(0)}_{\zeta,\eta}$, in addition, contains information of the perturbation (i.e., the magnetization).

The first order correction of the eigenfunction is calculated through the new basis,
\begin{align}
\Phi^{(1)}_{\zeta,\eta}  &=    \frac{ \Phi^{(0)\dagger}_{\zeta,-\eta} \bar{\mathcal{H}}_\zeta' \Phi^{(0)}_{\zeta,\eta} }{ E^{(1)}_{\zeta,\eta} - E^{(1)}_{\zeta,-\eta}  } \Phi^{(0)}_{\zeta,-\eta}   + \sum_{\eta'} \frac{ \Phi_{-\zeta,\eta'}^{(0)\dagger} \mathcal{H}'  \Phi^{(0)}_{\zeta,\eta} }{E^{(0)}_{\zeta,\eta'} - E^{(0)}_{-\zeta,\eta}}   \Phi^{(0)}_{-\zeta,\eta'} \notag \\
&= -\zeta M \frac{   k_{\parallel}    }{2  \tilde{k}^2 }  \Phi^{(0)}_{-\zeta,\eta}  ,
\label{eq:5_9}
\end{align}
where $\bar{\mathcal{H}}_\zeta' = P_\zeta \mathcal{H}' P_{-\zeta} (E_\zeta - \mathcal{H}_D)^{-1} P_{-\zeta} \mathcal{H}'  P_\zeta$ is the second order perturbation in the subspace $\mathbb{H}_{\zeta}$. 
Thus, up to the first order correction, the eigenenergy and eigenfunctions are
\begin{align}
\label{eq:5_10}
E_{\zeta, \eta} &= \zeta \tilde{k}+ \zeta \eta M \frac{\tilde{k}_\perp}{\tilde{k}} + \mathcal{O}(M^2) ,\\
\Psi^F_{\zeta,\eta} &= \Phi^{(0)}_{\zeta,\eta} 
 - \zeta M \frac{k_\parallel}{ \tilde{k}}  \Phi^{(0)}_{-\zeta,\eta}+ \mathcal{O}(M^2) ,
\label{eq:5_11}
\end{align}
which are fully consistent with the Taylor expansion of $E_{\zeta, \eta}$ and $\Psi^F_{\zeta,\eta} $ of the ferromagnetic Dirac Hamiltonian. Clearly, when the magnetization $(M)$ asymptotically approaches zero, the energy bands become spin degenerate [Eq.~(\ref{eq:5_10})], but the eigenfunctions return to $\Phi^{(0)}_{\zeta,\eta}$, which differs from $\Psi^D_{\zeta,\eta}$ by a specific gauge chosen by the direction of magnetization.

\subsection{SHC in spectral representation}
Next, we calculate the intrinsic SHC through the Kubo formula using the three sets of eigenfunctions, \textit{i.e.,} $\Psi^D_{\zeta,\eta},\Phi^{(0)}_{\zeta,\eta}$, and $ \Psi^F_{\zeta,\eta}$, to show the effect of different eigenfunctions on the SHC. 
The Kubo formula for SHC in the spectral representation \cite{guo2008intrinsic} is given by
\begin{align}
\label{eq:5_12}
\sigma^{l}_{ji} &= -\frac{e}{4} \sum_{\bm{k},\zeta,\eta} f (\varepsilon_{\bm{k}}) \Omega^{l,\zeta,\eta}_{ji},\\
\label{eq:5_13}
\Omega^{l,\zeta,\eta}_{ji} &= - \sum_{(\zeta',\eta') \neq (\zeta,\eta)} 2  \text{Im} \frac{ \phi_{\zeta,\eta}^{\dagger} v_j^l  \phi_{\zeta',\eta'}   \phi_{\zeta',\eta'}^{\dagger} v_i  \phi_{\zeta,\eta}  }{(E_{\zeta,\eta} -E_{\zeta',\eta'})^2 },
\end{align}
where $f (\varepsilon_{\bm{k}}) $ is the Fermi-Dirac distribution function and $\Omega^{l,\zeta,\eta}_{ji} $ are the Berry curvaturelike terms \cite{guo2008intrinsic,gradhand2012first} for band indices $\zeta,\eta$, where $(\zeta',\eta') \neq (\zeta,\eta)$ indicates summation over all interband transitions. In the following, we substitute $\Psi^D$, $\Phi^{(0)}$, and $\Psi^{F}$ into $\phi_{\zeta,\eta}$. Note that we calculate different components of the SHC tensor with a fixed magnetization direction: $\hat{m} \parallel \hat{z}$.

For the eigenfunctions of pristine Dirac Hamiltonian ($\Psi^D$), we consider all nonvanishing components of the totally antisymmetric SHC tensor. The corresponding Berry curvaturelike terms are
\begin{align}
\Omega^{D,3,\zeta,\eta}_{21} &=   \sum_{(\zeta',\eta') \neq (\zeta,\eta)}
  \frac{ \Delta  }{(\zeta-\zeta')\tilde{k}^3 }   \left( \frac{ k_1^2 +k_2^3}{k^2}  \delta_{\eta, \eta'} + \frac{ k_3^2}{k^2}  \delta_{\eta, -\eta'}  \right) \notag \\
 &= \zeta \frac{ \Delta  }{2 \tilde{k}^3},  
\label{eq:5_14}
\end{align}
where two transition channels $ \zeta,\eta \rightarrow -\zeta,\pm \eta $ equally contribute to $\Omega^{D,3,\zeta,\eta}_{21}$ due to the spin degeneracy. 
One may verify that $\Omega^{D,1,\zeta,\eta}_{32}$ and $\Omega^{D,2,\zeta,\eta}_{13}$ have similar expressions, which suggests that the $ \zeta,\eta \rightarrow -\zeta,\pm \eta $ channels contribute equally despite the geometry of SHC.

For the eigenfunctions $\Phi^{(0)}$, the corresponding Berry curvaturelike terms, denoted as $\Omega^{F_0,l,\zeta,\eta}_{ji}$,  are
\begin{align}
\label{eq:5_16_a} 
\Omega^{F_0,3,\zeta,\eta}_{21} &=   \Omega^{F_0,1,\zeta,\eta}_{32} = \sum_{(\zeta',\eta') \neq (\zeta,\eta)}   \frac{  \Delta  (\eta -\eta')    (\zeta \eta + \zeta' \eta')     }{ 2 ( \zeta  - \zeta'  )^2  \tilde{k}^3  } \notag \\
&=   \zeta \frac{ \Delta  }{2\tilde{k}^3 },  \\ 
  \Omega^{F_0,2,\zeta,\eta}_{13} &=    \sum_{(\zeta',\eta') \neq (\zeta,\eta)}   \frac{  \Delta  ( \eta + \eta') (\zeta \eta - \zeta' \eta') }{ 2  ( \zeta - \zeta'  )^2  \tilde{k}^3 } \notag \\
  &=  \zeta \frac{ \Delta }{2\tilde{k}^3 }.
\label{eq:5_16_b} 
\end{align}
As evident from the first line of Eq.~(\ref{eq:5_16_a}), transitions that conserve spin, i.e., $\eta^\prime = \eta$, vanish in the summation.
Thus, only spin-flipping transitions $\zeta, \eta \rightarrow - \zeta, -\eta $ contribute to $\Omega^{F_0,3,\zeta,\eta}_{21}$ and $\Omega^{F_0,1,\zeta,\eta}_{32}$. In contrast, for $\Omega^{F_0,2,\zeta,\eta}_{13}$, the transitions that flip the spin $\eta^\prime = - \eta$ vanishes as apparent from the second line of Eq.~(\ref{eq:5_16_b}).
Here, the spin-conserving transitions $\zeta, \eta \rightarrow -\zeta, \eta $ contribute to $\Omega^{F_0,2,\zeta,\eta}_{13}$.
As evident, Eqs.~(\ref{eq:5_16_a}) and (\ref{eq:5_16_b}) are identical to Eq.~(\ref{eq:5_14}).
This is because $\Phi^{(0)}$ only differs from $\Psi^D$ by a specific gauge chosen by the direction of magnetization.

For the eigenfunctions of Dirac ferromagnet $(\Psi^F_{\zeta,\eta})$, the corresponding Berry curvaturelike terms are as follows:
\begin{align}
\label{eq:5_15_a}
\Omega^{F,3,\zeta,\eta}_{21} &=   
 \Omega^{F,1,\zeta,\eta}_{32} \notag \\
 &=     \sum_{(\zeta',\eta') \neq (\zeta,\eta)}  \frac{   \Delta  (\eta -\eta')   (\zeta \eta E_{\eta'} Q_{\eta}  + \zeta' \eta' E_{\eta} Q_{\eta'}) }{ 2   \tilde{k}_\perp E_{\eta} E_{\eta'} ( \zeta E_\eta - \zeta'  E_{\eta'} )^2 } \notag \\
 &= \zeta \frac{ \Delta }{2E_\eta \tilde{k}_{\perp}^3   } \left[  \tilde{k}_{\perp}  + \frac{\eta}{M} ( k^2_\parallel+M^2)\right],  \\
\Omega^{F,2,\zeta,\eta}_{13}& =  \sum_{(\zeta',\eta') \neq (\zeta,\eta)}   \frac{  \Delta  (\eta + \eta')  (\zeta \eta E_{\eta'} Q_{\eta}  -  \zeta' \eta' E_{\eta} Q_{\eta'}) }{ 2  \tilde{k}_\perp E_{\eta} E_{\eta'} ( \zeta E_\eta - \zeta'  E_{\eta'} )^2 }\notag \\
&=   \zeta \frac{ \Delta }{2E_\eta^3  } \frac{Q_\eta}{\tilde{k}_{\perp} } .
\label{eq:5_15_b}
\end{align}
Only spin-flipping transitions $\zeta, \eta \rightarrow \pm \zeta, -\eta $ contribute to $\Omega^{F,3,\zeta,\eta}_{21}$ and $\Omega^{F,1,\zeta,\eta}_{32}$ and  the spin-conserving transitions $\zeta, \eta \rightarrow -\zeta, \eta $ contribute to $\Omega^{F,2,\zeta,\eta}_{13}$. Here, in the presence of the spin gap, $\Omega^{F,3,\zeta,\eta}_{21}$ and $\Omega^{F,1,\zeta,\eta}_{32}$ have an extra spin-flipping channel: $\zeta, \eta \rightarrow \zeta, -\eta $. As evident, $\Omega^{F_0,2,\zeta,\eta}_{13}$ is also equal to $\Omega^{F,2,\zeta,\eta}_{13}$ in the $M \to 0$ limit, but $\Omega^{F_0,3,\zeta,\eta}_{21}$ and $\Omega^{F_0,1,\zeta,\eta}_{32}$ do not take the same form with $\Omega^{F,3,\zeta,\eta}_{21}$ and $\Omega^{F,1,\zeta,\eta}_{32}$ in the $M \to 0$ limit due to the absence of the transition channel $\zeta, \eta \rightarrow \zeta, -\eta $ for the former (i.e., for the eigenstates $\Phi^{(0)}$). Furthermore, we note that the Berry curvaturelike terms all vanish with zero gap ($\Delta=0$). Thus, gap of Dirac cone $\Delta$ is essential for the nonzero SHE in the Dirac ferromagnet.

Since we have fixed the magnetization direction along the $z$ axis, $\Omega^{F,2,\zeta,\eta}_{13}$ corresponds to the case with $\hat{m} || \hat{E}$, whereas $\Omega^{F,3,\zeta,\eta}_{21}$ and $\Omega^{F,1,\zeta,\eta}_{32}$ represent the case with $\hat{m} \perp \hat{E}$.
%\textcolor{red}{
One can verify that Eq.~(\ref{eq:5_15_a}) is exactly the same with the integrand of Eq.~(\ref{eq:4_5}).
Thus, upon integrating the Berry curvaturelike terms across the $\bm{k}$ space, the SHC of the Dirac ferromagnet with $\hat{m} \perp \hat{E}$ is near zero. 
It turns out that the transition $\zeta, \eta \rightarrow \zeta, -\eta $ compensates most of the contribution from the transition $\zeta, \eta \rightarrow - \zeta, -\eta $, causing the small SHC with $\hat{m} \perp \hat{E}$.
Note that the compensation is exact in the asymptotic limit of zero magnetization.

The transition channels that contribute to the Berry curvaturelike terms are summarized in Table~\ref{Tab:2}.
With the degenerate perturbation theory, one can trace the origin of the discrepancy between pristine and ferromagnetic Dirac electron systems in the zero magnetization limit. 
In the pristine Dirac model, the spin space spanned by $\eta$ is degenerate, thus the nonvanishing components of the SHC tensor share the same transition channels ($ \zeta, \eta \rightarrow -\zeta, \pm \eta $). In the Dirac ferromagnet, however, the spontaneous magnetization $ \mathcal{H}'(M)$ breaks the spin degeneracy. Consequently, each nonvanishing component of the SHC tensor possesses distinct transition channels, either spin-flipping or spin-conserving channels (see Table~\ref{Tab:2}). Furthermore, in the presence of the spin gap, a unique spin-flipping transition channel $\zeta, \eta \rightarrow \zeta, -\eta $ emerges when $\hat{m} \perp \hat{E}$,
which nearly compensates contribution from the transition channel $\zeta, \eta \rightarrow -\zeta, -\eta $ on the SHC.
Thus, the appearance of the spin-flipping transition channel causes the SHC to be negligible when $\hat{m} \perp \hat{E}$. 
These results show that the change in the selection rule of the state transitions define the SHC and is the cause of the anisotropy in the SHC that persists to the $M \rightarrow 0$ limit. 

%We therefore attribute the difference of the SHC to the difference in the transition channels allowed for each phase.
%, across the paramagnetic to ferromagnetic transition.

\begin{table}
\caption{\label{Tab:2} Summary of transition channels in $\Omega^{l,\zeta,\eta}_{ji} $ for each eigenfunction set.}
\begin{ruledtabular}
\begin{tabular}{llll}
   & $\Psi_{\zeta, \eta }^D$ & $\Phi_{\zeta, \eta }^{(0)}$ & $\Psi_{\zeta, \eta }^F$ \\
  \hline
   $\hat{m} \parallel  \hat{E}$ &  $ \zeta, \eta \rightarrow -\zeta, \pm \eta $  & $\zeta, \eta \rightarrow -\zeta, \eta$  & $\zeta, \eta \rightarrow -\zeta, \eta$ \\
  $\hat{m} \perp \hat{E}$ & $ \zeta, \eta \rightarrow -\zeta, \pm \eta $ & $\zeta, \eta \rightarrow -\zeta, -\eta$ & $\zeta, \eta \rightarrow \pm \zeta, -\eta$ \\
\end{tabular}
\end{ruledtabular}
\end{table}

%%%%%%% Summary %%%%%%%%
\section{Summary}\label{Sec6:Summary}
To summarize, we have developed a minimal model to study the intrinsic SHE in a ferromagnetic Dirac system. 
We find the nonvanishing components of the SHC tensor is anisotropic.
The anisotropy is defined by the direction of magnetization. 
Specifically, the SHC is finite when the magnetization is parallel with the external electric field, whereas, it is near zero when the two are orthogonal.
Using the spectral representation of the Kubo formula, the nonvanishing components of the SHC tensor are examined in the context of transition channels.
We find the interband transition channels in the pristine Dirac Hamiltonian system bifurcate into spin-conserved or spin-flipping ones in the presence of the exchange field, which ultimately causes the anisotropy of SHC in the Dirac ferromagnet.
Interestingly, the anisotropy persists as the magnetization approaches zero.
The SHC is smoothly connected to the pristine Dirac phase when the magnetization is reduced to zero along the applied electric field.
In contrast, a discontinuity in the SHC appears at the ferromagnetic Dirac/pristine Dirac border when the magnetization is not parallel to the electric field. The discontinuity of the SHC at the ferromagnetic Dirac/pristine Dirac border is caused by the emergence of a spin-flipping transition channel spontaneously appearing with a spin gap in the Dirac ferromagnet.  

Our paper demonstrates the effect of ferromagnetic ordering on the intrinsic SHE using a minimal model combining ferromagnetism and SOC. The massive Dirac Hamiltonian has been employed as an effective Hamiltonian around the L point of semimetallic bismuth. A ferromagnetic Dirac system can be achieved by magnetic doping or the magnetic proximity effect from a neighboring layer. In the ferromagnetic Dirac phase, we placed a limit on the strength of spontaneous magnetization to be smaller than the gap of Dirac cone (i.e., $M < \Delta $). If the magnetization equals the gap, a gap closing occurs, which can be classified as a topological phase transition. In the extreme case where the magnetization is larger than the gap, the Dirac semimetal turns into a magnetic Weyl semimetal. Further theoretical study is required to address the SHE in the latter system.

\begin{acknowledgments}
This work was supported by Grants-in-Aid for Scientific Research (B) (Grants No. 21H01034 and No. JP18H01162) from the Japan Society for the Promotion of Science and by the JST-Mirai Program (Grant No. JPMJMI19A1).
%put your acknowledgments here.
\end{acknowledgments}

%\begin{widetext}
\appendix

\section{Wolff Hamiltonian}\label{Appendix:0}
In this appendix, we present the derivation of the Wolff Hamiltonian from the Kohn–Sham Hamiltonian and show its equivalence with the pristine Dirac Hamiltonian with  an isotropic velocity. We start with a general single electron Hamiltonian that includes the SOC,
\begin{align}
\hat{H} = \frac{\hat{p}^2}{2m} + V(\bm{r}) + \frac{\hbar}{ 4 m^2 c^2} \bm{\sigma} \cdot \nabla V(\bm{r}) \times \hat{\bm{p}} .
\label{eq:R1}
\end{align}
Following the $k\cdot p$ theory, the Hamiltonian can be expanded around a band extremum $\bm{k}_0$, \textit{e.g.}, the $L$ point of bismuth. Accordingly, the wave function around the extremum $\bm{k}_0$ can be constructed using the Bloch basis,
\begin{align}
\psi(\bm{r}) = \sum_{n} \int d \bm{k} c_n(\bm{k}) e^{i \bm{k} \cdot \bm{r}} u_{n,\bm{k}_0},
\label{eq:R2}
\end{align}
where $c_n(\bm{k})$ is an expansion coefficient and $u_{n,\bm{k}_0}$ is a function periodic in space. 
The matrix elements of the Hamiltonian around the  band extremum can be rewritten as
\begin{align}
\label{eq:R3}
\bra{n, \bm{k}}  \hat{H}  \ket{n', \bm{k}} c_{n'} (\bm{k}) &= E_n c_{n} (\bm{k}) \\
\bra{n, \bm{k}}  \hat{H}  \ket{n', \bm{k}} &= \left[ \varepsilon_{n,\bm{k}_0} + \frac{\hbar^2 k^2}{2m} \right] + \frac{\hbar{\bm{k}} \cdot \bm{p}_{n,n'}}{m},
\label{eq:R4}
\end{align}
where $\varepsilon_{n,\bm{k}_0}$ is the eigenenergy of the Bloch state at $\bm{k}_0$ and the momentum matrix elements $\bm{p}_{n,n'}$ are given as
\begin{align}
\bm{p}_{n,n'} = \frac{ (2\pi)^3}{\Omega} \int_{\text{cell}} d \bm{r} u^{*}_{n,\bm{k}_0} \left( \hat{\bm{p}} + \frac{\hbar}{4mc^2} \bm{\sigma} \times \nabla V(\bm{r}) \right) u_{n',\bm{k}_0}.
\label{eq:R5}
\end{align}
Equation~(\ref{eq:R5}) clearly shows that the SOC generates an anomalous velocity term in the momentum matrix element. Focusing on the low energy properties, we select the lowest conduction and highest valance bands (both spin degenerate, the band gap is $2\Delta$) \text to construct an effective Hamiltonian. Under such circumstance, Eq.~(\ref{eq:R4}) reduces to the Wolff Hamiltonian,
\begin{align}
\mathcal{H}=\hbar k_i W_i^{j} \rho_{2} \otimes \sigma^j + \Delta \rho_3 \otimes \sigma^0,
\label{eq:R6}
\end{align}
with the Wolff tensor $W_i^{1} = \text{Im} (p_{i,1,4}/m)=-\text{Im} (p_{i,3,2}/m) , W_i^{2} = \text{Re} (p_{i,1,4}/m)=-\text{Re} (p_{i,3,2}/m)$, and $W_i^{3} = \text{Im} (p_{i,1,3}/m)=\text{Im} (p_{i,4,2}/m)$. The band indices $1,2$ and $3,4$ are the spin degenerate conduction bands and the valance bands, respectively. If the Wolff tensor is approximately isotropic, i.e.,  $W_i^{j}= \delta_{ij}  v$, Eq.~(\ref{eq:R6}) is equivalent to Eq.~(\ref{eq:2_1}) without magnetization. We note that the representation of the velocity operator $(\rho_{2} \otimes \bm{\sigma})$ in the Wolff Hamiltonian differs from the original Dirac Hamiltonian up to a unitary transformation. 

The exchange coupling term in Eq.~(\ref{eq:2_1}) is obtained using the spin magnetic moment operator \cite{fuseya2015transport}.  To derive the spin magnetic moment operator in the Dirac Hamiltonian, we first rewrite the wavevector as the momentum operator in the pristine Dirac Hamiltonian and perform Peierls substitution $\hbar \bm{k} \rightarrow  \hat{\bm{p}}  \rightarrow  \hat{\bm{\pi}} = \hat{\bm{p}} + e/c \bm{A}$,
\begin{align}
\mathcal{H}_0= v \bm{\pi}_i \rho_1 \otimes \sigma^i +  \Delta \rho_3 \otimes \sigma^0 .
\label{eq:R7}
\end{align}
Applying Foldy and Wouthuysen transformation \cite{foldy1950dirac} generated by $S =  \frac{v}{2  \Delta }    \rho_2  \otimes \sigma^i  \hat{\pi}_i  $, the Dirac Hamiltonian can be decoupled into positive and negative energy states,
\begin{align}
\mathcal{H}_0' &= e^{i S} \mathcal{H}_0 e^{-i S} \notag \\
&= \mathcal{H}_0 + i \left[ S ,  \mathcal{H}_0 \right] + \frac{i^2}{2!}  \left[ S , \left[ S ,  \mathcal{H}_0 \right]  \right] + \cdots \notag \\
&= ( \Delta  + \frac{v^2 \bm{\pi}^2}{2 \Delta}  )\rho_3 \otimes \sigma^0  +  \frac{ \hbar e v^2 }{ 2 \Delta c}  B_i  \rho_3  \otimes \sigma^i    + \mathcal{O}(v^2/\Delta^2)
\label{eq:R8}
\end{align}
where the external magnetic field arises from the vector potential $\bm{B} = \nabla \times \bm{A}$. The spin magnetic operator $\rho_3  \otimes \sigma^i $ couples to external magnetic field oppositely in positive (electrons) and negative (holes) energy states.

\section{Intrinsic SHC in the Kubo formula}\label{Appendix:A}
The intrinsic SHC is obtained by calculating the Fermi surface and Fermi sea terms with zeroth order of damping constant $\mathcal{O}(\gamma^0)$.
\subsection{ Fermi surface contribution}
Substituting $\tilde{G}^{R,A}$ into Eq.~(\ref{eq:3_4}), the Fermi surface term reads
\begin{align}
&\sigma_{ij}^{k,(1)} =   \frac{e}{ 8 \pi V} \sum_{\bm{k}}  \epsilon_{ik l}
tr\left(  \rho_{\mu} \rho_2 \rho_{\lambda} \rho_1 \right) tr\left(  \sigma^{\nu} \sigma^{l} \sigma^{\tau} \sigma^{j}  \right) \nonumber \\
& \times  \left(   \frac{ g^{R}_{\mu\nu}(\varepsilon)  g^{R}_{\lambda \tau}(\varepsilon) }{D^R(\varepsilon)D^R(\varepsilon)} -2 \frac{ g^{A}_{\mu\nu}(\varepsilon)  g^{R}_{\lambda \tau}(\varepsilon) }{D^A(\varepsilon)D^R(\varepsilon)} + \frac{ g^{A}_{\mu\nu}(\varepsilon)  g^{A}_{\lambda \tau}(\varepsilon) }{D^A(\varepsilon)D^A(\varepsilon)} \right) \Big|_{\varepsilon=\mu}  .
 \label{eq:3_6}
\end{align}

The numerator and denominator of Eq.~(\ref{eq:3_6}) can be expanded on the order of $\gamma$,
\begin{eqnarray}
g^{R,A} (\varepsilon) &=&g (\varepsilon) \pm i \gamma g' (\varepsilon) +\mathcal{O} ( \gamma^2), \notag \\
D^{R,A} (\varepsilon) &=&D (\varepsilon) \pm i \gamma D' (\varepsilon) +\mathcal{O} ( \gamma^2).
 \label{eq:3_9}
\end{eqnarray}
Note that for the denominators, we have the following approximation:
\begin{align}
  \label{eq:3_10}
 \frac{1}{  D^2(\varepsilon)  + \gamma^2 (D'(\varepsilon))^2}  & \simeq \frac{\pi}{ |D'(\varepsilon) | \gamma} \delta(D(\varepsilon)) .
\end{align}
 Taking the zeroth order of $\gamma$, the Fermi surface term reads
\begin{eqnarray}
\sigma_{21}^{3,(1)} &=&   \frac{ 2 e}{  V} \sum_{\bm{k}}  \frac{ \delta(D(\varepsilon)) }{|D'(\varepsilon)|}  X^{(0)}(\varepsilon) \Big|_{\varepsilon=\mu}.
 \label{eq:3_11}
\end{eqnarray}

\subsection{ Fermi sea contribution}
Substituting $\tilde{G}^{R,A}$ into Eq.~(\ref{eq:3_5}), the Fermi sea term reads
\begin{align}
&\sigma_{ij}^{k,(2)}
=  \frac{e}{ 8 \pi V} \sum_{\bm{k}}  \epsilon_{ik l} tr\left(  \rho_{\mu} \rho_2 \rho_{\lambda} \rho_1 \right) tr\left(  \sigma^{\nu} \sigma^{l} \sigma^{\tau} \sigma^{j}  \right) \notag \\
& \times  \int^{ \mu}_{-\infty} d\varepsilon  
 \left[ \frac{ g^{R}_{\mu\nu}(\varepsilon)\partial_z  g^{R}_{\lambda \tau}(\varepsilon) - \partial_\varepsilon g^{R}_{\mu\nu}(\varepsilon)  g^{R}_{\lambda \tau}(\varepsilon) }{ (D^R(\varepsilon) )^2}    - \Big( R \leftrightarrow A\Big)  \right]   .
\label{eq:3_13}
\end{align}
For Eq.~(\ref{eq:3_13}), the tensor component $\sigma_{21}^{3,(2)}$ reads
\begin{align}
&\sigma_{21}^{3,(2)}
= - \frac{2e}{  \pi V} \sum_{\bm{k}} \int^{ \mu}_{-\infty} d\varepsilon \   \text{Im} \Big{[}  \frac{ X^R(\varepsilon) }{ D^{R}(\varepsilon)^2} \Big{]} .
\label{eq:3_14}
\end{align}
Taking the zeroth order of $\gamma$, the integrand in Eq.~(\ref{eq:3_14}) is approximated as 
\begin{eqnarray}
 &&\text{Im} \Big{[}  \frac{ X^R(\varepsilon) }{ D^{R}(\varepsilon)^2} \Big{]}   \approx \pi  \ \text{sgn} D'( \zeta E_{\eta})  \partial_\varepsilon  \delta(D(\varepsilon)) \frac{X^{(0)}(\varepsilon)}{D'(\varepsilon)} .
\label{eq:3_16}
\end{eqnarray}
where $X^{R}(\varepsilon)$ is expanded on the order of $\gamma$,
\begin{eqnarray}
X^{R}(\varepsilon) = X^{(0)}(\varepsilon) + i \gamma X'^{(0)}(\varepsilon) + \mathcal{O} ( \gamma^2).
\label{eq:3_15}
\end{eqnarray}
The intrinsic Fermi sea term thus reads
\begin{align}
\sigma_{21}^{3,(2)}&=
 - \frac{ 2 e }{  V} \sum_{\bm{k}}   \int^{ \mu}_{-\infty} d\varepsilon  \ \text{sgn} D'( \zeta E_{\eta})   \partial_\varepsilon  \delta(D(\varepsilon))  \frac{X^{(0)}(\varepsilon)}{D'(\varepsilon)} \nonumber  \\
&=- \frac{ 2 e }{  V} \sum_{\bm{k}} \Big[  \text{sgn} D'( \zeta E_{\eta}) \frac{\delta(D(\varepsilon))}{D'(\varepsilon)}  X^{(0)}(\varepsilon) \Big|^{\varepsilon = \mu}_{\varepsilon = -\infty} \notag \\
& -   \int^{ \mu}_{-\infty} d\varepsilon \ \text{sgn} D'( \zeta E_{\eta})  \delta(D(\varepsilon))\partial_\varepsilon \Big( \frac{X^{(0)}(\varepsilon)}{D'(\varepsilon)} \Big) \Big],
\label{eq:3_17}
\end{align}
where we apply integration by parts in the last equation. 
Note that the first term  in Eq.~(\ref{eq:3_17}) with the lower boundary vanishes at $\varepsilon \to - \infty$ and its upper boundary term exactly cancels the Fermi surface term [Eq.~(\ref{eq:3_11})]. Consequently, the total intrinsic SHC reduces to a compact form shown in Eq.~(\ref{eq:3_18}).

\subsection{Anisotropic SHC}
To obtain an analytical expression of Eq.~(\ref{eq:3_18}), we first rewrite $\delta(D(\varepsilon))$,
\begin{eqnarray}
\delta(D(\varepsilon)) &=& \sum_{\eta,\zeta = \pm1} \frac{\delta(\varepsilon- \zeta E_{\eta})}{8 M E_{\eta} \tilde{k}_{\perp} } ,
\label{eq:4_1}
\end{eqnarray}
where we denote $ \tilde{k}_{\perp} = \sqrt{\Delta ^2 + k_{\perp}^2}$. Hence, Eq.~(\ref{eq:3_18}) is rewritten as
\begin{align}
\sigma_{21}^{3} &=  \frac{ 2e}{  V} \sum_{\bm{k},\eta,\zeta}   \int^{ \mu}_{-\infty} d\varepsilon\    \frac{\delta(\varepsilon- \zeta E_{\eta})}{64M^2 E_{\eta}^2  \tilde{k}_{\perp}^2 } \notag \\
 &\times  \Big[    \partial_\varepsilon  X^{(0)}(\varepsilon)   -   \zeta X^{(0) }(\varepsilon)  (\frac{1}{E_{\eta}} +  \frac{\eta}{M} \frac{E_{\eta}}{  \tilde{k}_{\perp} } )  \Big] .
\label{eq:4_2}
\end{align}
Owing to  $\delta(\varepsilon- \zeta E_{\eta})$ in the integral, we replace $\varepsilon$ in $X^{(0)}(\varepsilon)$ and $X'^{(0)}(\varepsilon)$ with $\zeta E_{\eta}$, which gives
\begin{align}
X^{(0)}(\zeta E_{\eta}) &= - 8 \Delta  M^2 \left[  { (1-m_1^2) E_{\eta}}^2  +  m_1^2 \eta  M \tilde{k}_{\perp}  + m_1^2 \tilde{k}_{\perp}^2  \right], \notag \\
\label{eq:4_3}
X'^{(0)}(\zeta E_{\eta}) &=  -8 \Delta  \zeta E_{\eta} \left[  2 M^2 + \eta  M \tilde{k}_{\perp}  -  M_1^2     \right] .
\end{align}

Equation~(\ref{eq:4_2}), thus, reads
\begin{align}
\sigma_{21}^{3}&= - \frac{ e \Delta  }{ 4 V} \sum_{\bm{k},\eta,\zeta}   \int^{ \mu}_{-\infty} d\epsilon    \delta(\varepsilon- \zeta E_{\eta}) \Big[ m_1^2  \left( - \frac{  \eta  M   }{ E_{\eta}^3   \tilde{k}_{\perp} }  -   \frac{ 1}{ E_{\eta}^3  } \right)  \notag \\
&+  \left(   \frac{  1 }{ {E_{\eta}}  \tilde{k}_{\perp}^2 }    +  \frac{ \eta  }{ M } \frac{1 }{ {E_{\eta}}  \tilde{k}_{\perp} }   -   \frac{ \eta  }{ M }  \frac{ {E_{\eta}}}{   \tilde{k}_{\perp}^3 } \right) \Big],
\label{eq:4_3_2}
\end{align}
from which we obtain Eqs.~(\ref{eq:4_5}) and (\ref{eq:4_6}) by separating the isotropic- and magnetization-dependent parts.

\subsection{Calculation of $\sigma^{3,\mathrm{iso}}_{21}$ and $\sigma^{3,\hat{m}}_{21}$}
$\sigma^{3,\mathrm{iso}}_{21}$ and $\sigma^{3,\hat{m}}_{21}$ both contain $\varepsilon$ integrals with an integration limit that approaches minus infinity $(\varepsilon \to - \infty)$.
One needs to introduce a proper energy cutoff to avoid the ultraviolet divergence in the effective Dirac Hamiltonian\cite{fujimoto2013ultraviolet,takane2019gauge}. We, therefore, separate the $\varepsilon$ integrals in $\sigma^{3,\mathrm{iso}}_{21}$ and $\sigma^{3,\hat{m}}_{21}$ into a divergent part $\left[\varepsilon \in(-\infty,0) \right]$ and a convergent part $\left[ \varepsilon \in(0,\mu) \right]$,
\begin{align}
\label{eq:4_7}
\sigma^{3,\mathrm{iso}}_{21} &= D^{\mathrm{iso}}+ C^{\mathrm{iso}}, \\
\sigma^{3,\hat{m}}_{21} &= D^{\hat{m}}+ C^{\hat{m}},
\label{eq:4_8}
\end{align}
where $C^{\mathrm{iso}}$, $C^{\hat{m}}$ and $D^{\mathrm{iso}}$, $D^{\hat{m}}$ are the convergent and divergent terms of the isotropic and the $\hat{m}$-dependent contributions, respectively.

The divergent terms ($D^{\mathrm{iso}}$ and $D^{\hat{m}}$) represent the SHC with the chemical potential exactly placed at the center of the gap between the positive and the negative energy states ($\mu=0$). In other words, it contains contribution from the entire negative energy states $(\zeta=-1,\eta= \pm1)$,
\begin{align}
\label{eq:4_9}
D^{\mathrm{iso}}  &=  \frac{ e \Delta  }{ 4 V} \sum_{\bm{k},\eta}   \frac{  1 }{ {E_{\eta}}  \tilde{k}_{\perp}^2 }     +  \frac{ \eta  }{ M } \frac{1 }{ {E_{\eta}}  \tilde{k}_{\perp} }   -   \frac{ \eta  }{ M }  \frac{ {E_{\eta}}}{   \tilde{k}_{\perp}^3 }  , \\
 \label{eq:4_10} 
D^{\hat{m}}
 &=- \frac{e  \Delta  }{ 4 V} \sum_{\bm{k},\eta}     \frac{  \eta  M   }{   \tilde{k}_{\perp} E_{\eta}^3 }  +   \frac{1}{ E_{\eta}^3 }.
\end{align}
As the $\bm{k}$ summation involves a logarithmic divergence in three dimensional reciprocal space, we adopt a cutoff energy $\Lambda_{\varepsilon}$. $\Lambda_{\varepsilon}$ is defined such that the following condition is satisfied:
\begin{equation}
E^2_{\eta } = k_{\parallel}^2 +(\tilde{k}_{\perp} +\eta M)^2  \leqslant \Lambda_\epsilon^2.
\label{eq:4_15}
\end{equation}
With the energy cutoff, the spin-split bands in the negative-energy states are truncated by two different ellipsoids [see Fig.~\ref{fig:1}(b)]. Setting $\tilde{M} = M/\Delta$, we expand $\tilde{M}$ up to second order in $\tilde{M}$ to obtain an analytical expression for $D^{\mathrm{iso}}$ and $D^{\hat{m}}$ ($M/\Delta $ is placed back in),
\begin{align}
\label{eq:4_16}
D_{\Lambda_\varepsilon}^{\mathrm{iso}} &=   \frac{ e \Delta  }{ 4\pi ^2}    \frac{M^2}{ 6 \Delta ^2} +\mathcal{O} (\tilde{M}^4), \\
\label{eq:4_17}
D_{\Lambda_\varepsilon}^{\hat{m}} &= - \frac{  e \Delta  }{ 4\pi^2} \left(   \ln   \frac{2\Lambda_{\varepsilon}}{\Delta } - 1  + \frac{M^2}{2\Delta ^2} \right) +\mathcal{O} (\tilde{M}^4).
\end{align}

For the convergent terms ($C^{\mathrm{iso}}$ and $C^{\hat{m}}$), the $\bm{k}$ integration is confined within the area that satisfies the following condition,
\begin{eqnarray}
E_{\eta}^2 = k_{\parallel}^2 + (\tilde{k}_{\perp} +\eta M)^2 \leqslant \mu^2.
\label{eq:4_19}
\end{eqnarray}
The integration range is similar to that of the energy cutoff scheme.
Again, the energy states are truncated by two different ellipsoids associated with the spin-up and spin-down bands $(\eta= \pm1)$.
Here, either the positive- or the negative-energy state $(\zeta=\pm1)$ is truncated. After some calculations, $C^{\mathrm{iso}}$ and  $C^{\hat{m}}$ read
\begin{align}
C^{\mathrm{iso}} &=  \frac{ e \Delta  }{  8 \pi^2 }  \sum_{\eta=\pm1}   \frac{ \eta   |\mu|  }{ 2M}   \Theta (\mu) \int_{\theta_{\alpha}}^{ \theta_{\beta}}  d\theta      \frac{ \cos^2 \theta }{  ( \sin \theta -\eta \frac{M}{|\mu|} )^2  } \notag\\
\label{eq:4_20}
&- \Big( 1  - \frac{ M^2  }{ |\mu|^2 ( \sin \theta -\eta \frac{M}{|\mu|} )^2  } \Big) \cos \theta \tanh^{-1} \cos \theta   \\
 C^{\hat{m}}&= -\frac{ e \Delta  }{ 8\pi^2 }  \sum_{\eta=\pm1}  \Theta (\mu) ( \cos \theta_{\alpha} + \ln \tan \frac{\theta_{\alpha}}{2} )
 \label{eq:4_21}
\end{align}
where $\theta_{\alpha} = \sin^{-1} \frac{ \Delta +\eta M}{|\mu|}$, $\theta_{\alpha}= \frac{\pi}{2}$, and $ \Theta (\mu) $ is defined as
\begin{eqnarray}
\Theta_{\eta} (\mu) =\left\{
\begin{aligned}
&1, & |\mu| > \Delta -\eta M \\
&0, &(\text{otherwise})
\end{aligned}
\right.
\label{eq:4_22}
\end{eqnarray}

Note that $\sigma^3_{21}$ is artificially divided into the convergent part and the divergent part. Hence, the convergent part is expected to exactly cancel the divergent part when the chemical potential approaches negative infinity. This is indeed the case. The cancellation between the $C$ terms with $\mu=\Lambda_{\varepsilon}$ and the $D$ terms justifies the energy cutoff scheme.

\subsection{Cutoff schemes for ultraviolet divergence}
In the pristine Dirac phase, a momentum cutoff scheme is typically employed to deal with the ultraviolet divergence.
In such scheme, the wave-vector $\bm{k}$ is truncated within a sphere of radius $\Lambda$ in the momentum space,
\begin{equation}
k_{\parallel}^2+ k_{\perp}^2 \leqslant \Lambda^2.
\label{eq:4_11}
\end{equation}
Note that, in the Dirac ferromagnet, the Fermi sea of spin-split bands $(\eta = \pm1) $ form ellipsoids with axial anisotropy along $k_{\parallel}$. The ellipticity of the Fermi sea asymptotically approaches 1 when $|\bm{k}| \gg M$ and the Fermi sea of the spin-split bands tend to coincide at the cutoff momentum $\Lambda \gg M$. In the momentum cutoff scheme, $D_{\Lambda}^{\mathrm{iso}}$ and $D_{\Lambda}^{\hat{m}}$ are approximated in the small magnetization limit as
\begin{align}
\label{eq:4_12}
D_{\Lambda}^{\mathrm{iso}} &=  - \frac{ e \Delta  }{ 4\pi ^2} \left( \ln  \frac{2 \Lambda}{ \Delta }  -1 -\frac{M^2}{ 6 \Delta ^2} \right) + \mathcal{O} (\tilde{M}^4), \\
\label{eq:4_13}
D_{\Lambda}^{\hat{m}} &= - \frac{  e \Delta  }{ 4\pi^2} \left(   \ln    \frac{2 \Lambda}{\Delta }  - 1  + \frac{M^2}{2 \Delta ^2} \right) + \mathcal{O} (\tilde{M}^4) .
\end{align}
 $D^{\hat{m}}$ shows exactly the same result for two cutoff schemes, whereas $D^{\mathrm{iso}}$ is different for the two. 
For example, in the limit of $M\rightarrow 0$, $D^{\mathrm{iso}}$ vanishes in the energy cutoff scheme but is finite for the momentum cutoff scheme. 
The difference is due to the ellipticity of Fermi sea of the negative energy branches, causing a gap between the Fermi surface of the two spin-split bands [see Fig.~(\ref{fig:1}b)] whose area is on the order of $M/\Lambda$. 
$D^{\mathrm{iso}}$ contains terms on the order of $M^{-1}$ [Eq.~(\ref{eq:4_9})], which return a nonvanishing contribution to the momentum integration even in the gap area. 
Counterintuitively, the small gap between the spin-splitting Fermi surfaces is significant to the whole Fermi sea term and does not vanish in the limit of $M\rightarrow 0$.

\begin{figure}[h]
 \includegraphics[width=8cm]{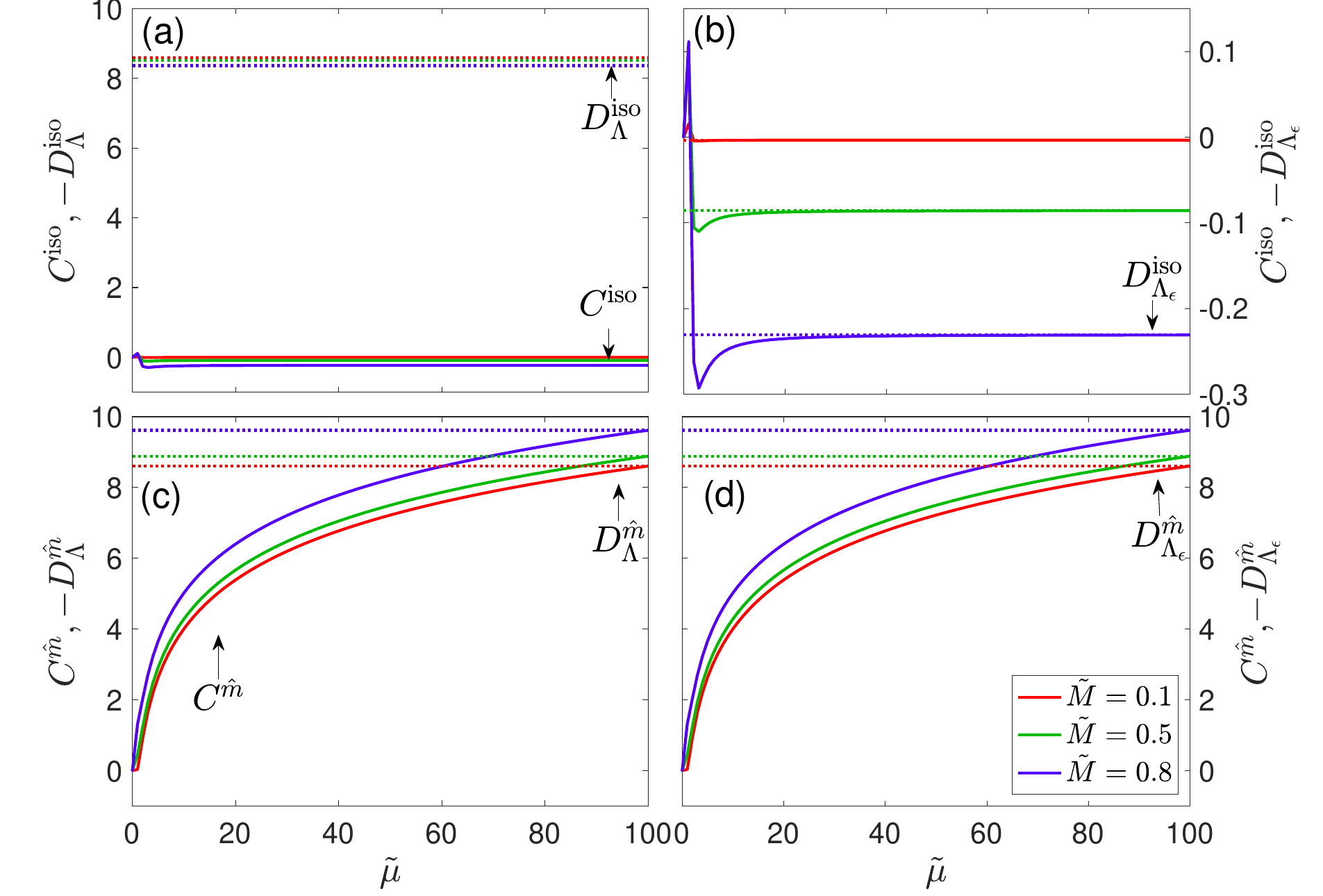}
 \caption{\label{fig:5}  Comparison between (a) $C^{\mathrm{iso}}$ and $-D^{\mathrm{iso}}_{\Lambda}$ (dashed) ;  (b) $C^{\mathrm{iso}}$ and $-D^{\mathrm{iso}}_{{\Lambda}_{\varepsilon}}$(dashed); (c) $C^{\hat{m}}$ and $-D^{\hat{m}}_{\Lambda}$ (dashed); (d) $C^{\hat{m}}$ and $-D^{\hat{m}}_{{\Lambda}_{\varepsilon}}$ (dashed). $M$ is set to be $0.1, 0.5, 0.8$, corresponding to color sequence from red to purple. The cutoff momentum/energy is set to be $\Lambda=\Lambda_\varepsilon=100$, according to the limit of chemical potential $\tilde{\mu}_{min}=-100$. The divergent parts are taken in opposite sign for comparison. The conductivity is in the unit of $\sigma^0_{s}=e \Delta /8\pi^2$.}
 \end{figure}

Additionally, we compare the convergent part and the divergent part to check the consistency of two cutoff schemes. 
For the momentum cutoff, the convergent part and the divergent part are different for the isotropic contributions $C^{\mathrm{iso}}$ and $D^{\mathrm{iso}}_{\Lambda}$ [see Fig.~\ref{fig:5} (a)], which do not reach the same value with $\mu \to \Lambda$.
In contrast, for the energy cutoff scheme, the convergent part and the divergent part are consistent for both the isotropic contribution [see Fig.~\ref{fig:5} (c)] and  the anisotropic contribution [see Fig.~\ref{fig:5} (d)]. 
Note that the convergent part should exactly cancel the divergent part at infinite negative energy $(-\infty \textrm{ or } \Lambda)$, since no states contributes to SHE. 
This indicates that the energy cutoff scheme should be the appropriate approach.
%%%%%%%%%% fig:5 %%%%%%%%% 
 %%%%%%%%%%%  fig:5 %%%%%%%%% 

 \section{General magnetization direction}\label{Appendix:B}
 \setcounter{equation}{0}
\numberwithin{equation}{section}
To generalize the discussion with an arbitrary magnetization direction [note that we chose $\bm{M} \parallel \hat{z}$ in $\mathcal{H}_F$, see Eq.~(\ref{eq:5_3})], we define a global unitary transformation which connects the two Hamiltonians $ \mathcal{H}_0$ and $\mathcal{H}_F$,
\begin{align}
U^\dagger \mathcal{H}_0 U = \mathcal{H}_F,
\label{eq:5_18}
\end{align}
with unitary matrix $ U=u_i \rho_3 \otimes \sigma^i $ and $\bm{u}= ( \sin \theta/2 \cos \varphi,  \sin \theta/2 \sin \varphi , \cos \theta/2 )$. Note that we have defined $\bm{M} = M (\sin\theta \sin\varphi, \sin\theta \sin\varphi, \cos\theta)$. The corresponding eigenfunctions $\Psi^{F'}_{\zeta, \eta}$ with such $\bm{M}$ read
\begin{align}
\mathcal{H}_0 \Psi^{F'}_{\zeta, \eta} = E_{\zeta, \eta} \Psi^{F'}_{\zeta, \eta}, \quad
\Psi^{F'}_{\zeta, \eta} = U \Psi^F_{\zeta, \eta}.
\label{eq:5_19}
\end{align}
Substituting $\Psi^{F'}_{\zeta, \eta}$ into the Kubo formula [Eq.~(\ref{eq:5_13})], the Berry curvaturelike terms take the following form:
\begin{align}
(\Omega^l_{ji})' &= \Omega^l_{ji} -  2 u_m u_k  \epsilon_{jlk}  \epsilon_{abm} \Omega^b_{ai} -2  u_i u_m   \Omega^l_{jm} \notag \\
&+ 4 u_m u_k u_i u_n \epsilon_{jlk}   \epsilon_{abm} \Omega^b_{an},
\label{eq:5_20}
\end{align}
Choosing the component $\Omega^3_{21}$ as an example, we obtain
\begin{align}
(\Omega^{F,3,\zeta,\eta}_{21} )'
&= \Omega^{F,3,\zeta,\eta}_{21} +  m^2_1 ( \Omega^{F,2,\zeta,\eta}_{13}  -  \Omega^{F,3,\zeta,\eta}_{21}) ,
\label{eq:5_21}
\end{align}
which takes a similar form to that shown in Eq.~(\ref{eq:4_4}). For example, when the magnetization is directed along the $x$ axis $(m_1=\pm1)$, Eq.~(\ref{eq:5_21})  satisfies the relation $(\Omega^{F,3,\zeta,\eta}_{21} )' =   \Omega^{F,2,\zeta,\eta}_{13}$, which indicates that spin-conserving channel contributes to the SHC. On the other hand, if the magnetization points orthogonal to the $x$ axis $(m_1=0)$, Eq.~(\ref{eq:5_21}) shows $(\Omega^{F,3,\zeta,\eta}_{21} )'
= \Omega^{F,3,\zeta,\eta}_{21} $ where only the spin-flipping channels are allowed.

\bibliography{Dirac_Ferro_072723}% Produces the bibliography via BibTeX.

\end{document}